\documentclass[11 pt,oneside,onecolumn,a4paper]{article}
\usepackage{epsfig,graphicx}
\usepackage{cite}
\usepackage{indentfirst}
\usepackage{amsfonts,amssymb,latexsym,amsmath,enumerate,amsthm}
\usepackage{indentfirst,cite}
\usepackage{bm}
\usepackage{amsmath}
\usepackage{amsmath}
\newcommand{\angstrom}{\textup{\AA}}
\def\Tr{{\rm Tr}}

\def\Im{{\rm Im}}
\def\Re{{\rm Re}}

\def\diag{{\rm diag}}
\def\det{{\rm det}}

mm \leftmargin 5pt \rightmargin 5pt \hoffset= -15mm

\title{Scattering of wave packets with phases} 

\author{Dmitry V.~Karlovets}

\date{\small{Department of Physics, Tomsk State University, \\ Lenina Ave. 36, 634050 Tomsk, Russia}}

\begin{document}

\maketitle

\begin{abstract}

A general problem of $2\rightarrow N_f$ scattering is addressed 
with all the states being wave packets with arbitrary phases.
Depending on these phases, one deals with coherent states in $(3+1)$ D, vortex particles with orbital angular momentum, 
the Airy beams, and their generalizations. A method is developed in which a number of events represents a functional of the Wigner functions
of such states. Using width of a packet $\sigma_p/\langle p\rangle$ as a small parameter, 
the Wigner functions, the number of events, and a cross section are represented
as power series in this parameter, the first non-vanishing corrections to their plane-wave expressions are derived,
and generalizations for beams are made. Although in this regime the Wigner functions turn out to be everywhere positive, 
the cross section develops new specifically quantum features, inaccessible in the plane-wave approximation. 
Among them is dependence on an impact parameter between the beams, on phases of the incoming states, 
and on a phase of the scattering amplitude. A model-independent analysis of these effects is made.
Two ways of measuring how a Coulomb phase and a hadronic one change with a transferred momentum $t$ are discussed.


\end{abstract}


\section{Introduction}

\subsection{Non-plane-wave states}

In a quantum theory of scattering, the in/out-states are commonly chosen as delocalized plane waves. 
This model allows one to tremendously simplify the calculations; however, 
its limits of applicability are not explicitly articulated in the overwhelming majority of textbooks.
Under standard conditions, finite sizes of the wave packets, their spreading during a collision, 
and finiteness of an interaction region do not play any essential role, especially for ultrarelativistic energies. 
There are, however, important exceptions. 

The first example in which this model fails to work is collision of beams with the large impact parameters -- 
the so-called MD-effect, observed at the collider VEPP-$4$ in Novosibirsk \cite{Impact}. 
A somewhat similar effect is collision processes with a $t$-channel singularity when initial particles are unstable \cite{t1, t2}.
It is beams' finite sizes that provide natural regularization of the divergence.  
The next illustration is neutrino oscillations -- a phenomenon that is intrinsically spatially and temporarily localized \cite{Akhmedov_09, Akhmedov_10, Akhmedov_Found}.

If colliding particles' wave fronts are neither plane nor gaussian, even approximately, then the plane-wave approximation is no longer valid either. 
The so-called vortex (or twisted) particles with orbital angular momentum (OAM) 
relative to a propagation axis and the Airy beams represent the simplest examples of such non-plane-wave states.
They were shown to be solutions of the wave equations \cite{Airy, Allen, BB, Bagrov, Airy_beam, Bliokh_07, Bliokh_11, Mono}, 
and the corresponding beams of photons, electrons, and neutrons were generated in recent years \cite{Uchida, Verbeeck, McMorran, Airy_Exp, Airy_El_Exp, neu}.
Vortex electrons with the kinetic energy of $200-300$ keV can be focused to a spot of an $\text{\angstrom}$ngstr$\ddot{\text{o}}$m size \cite{Angstrom}, 
their OAM can be as high as $\ell = 200\hbar$ \cite{Grillo15}, their magnetic moment increases proportionally to $\ell$ \cite{Bliokh_11}, 
and this brings about new effects in the electromagnetic radiation \cite{PRL}. Such photons and electrons were also proved useful for optical manipulation \cite{Mono}, 
for probing phase of a transition amplitude, for creating pairs entangled in their OAM \cite{Serbo, I_PRD, I_S, I_Phase, Ivanov_PRA_2012}, etc.

Besides that, several groups have recently managed to create even more sophisticated photonic quantum states, 
including those combining features of the vortex- and Airy beams \cite{AB_2010, Zhao, Mathieu};
and one can await generation of the corresponding states of massive leptons and hadrons in near future. 
Although these novel beams differ from the coherent states, as they possess a distinct set of quantum numbers, 
the difference between them, mathematically speaking, lies only in the phases of their wave functions $\psi ({\bm p})$.

A consistent relativistic scattering theory beyond the plane-wave approximation is absent by now,
even though a number of non-plane-wave solutions for relativistic wave equations have long been known \cite{Bagrov_Mono} 
and specific calculations were made \cite{Impact, t1, t2, Akhmedov_09, Akhmedov_10, Akhmedov_Found}.
For vortex beams, a corresponding ad hoc formalism has been recently developed by Ivanov and Serbo \cite{I_PRD, I_S}.
Generalization of their procedure for other quantum states (e.g. for coherent states, the Airy beams, etc.) may nevertheless represent a challenge. 
It is highly desirable therefore to have at hand an approach that would enable us to study scattering of the wave packets 
\textit{whatever their wavefront is}, that is, for arbitrary phases of their wave functions. 
This work aims at developing such a method by generalizing the customary (plane-wave) S-matrix formalism.
In doing so, we follow the work \cite{Impact} in which the incoming particles are described by their Wigner functions.


\subsection{Why Wigner functions?}

The reader may well ask why we should deal with the Wigner formalism,
especially when we are already going to treat sophisticated non-plane-wave effects. 
Indeed, although this approach was successfully applied both for non-relativistic- and relativistic scattering problems (see \cite{Caruthers} and \cite{Impact}, respectively),
it does not seem to have drawn much attention. The answer is that this formalism turns out to be the most convenient and elegant tool 
in the paraxial regime when $\sigma_p \ll \langle p \rangle$,
complementary and alternative to the wave-function approach 
(cf. \cite{Bagrov_TCS, Akhmedov_10, Akhmedov_Found}).

To put it in more detail, consider a generic matrix element 
\begin{eqnarray}
& \displaystyle
S_{fi} = \prod\limits_{i,f} \int \frac{d^3 p_i}{(2\pi)^3}\frac{d^3 p_f}{(2\pi)^3}\,\, \psi_i ({\bm p}_i) S_{fi}^{(pw)} \psi_f^* ({\bm p}_f)
\label{S}
\end{eqnarray}
which is a functional of the plane-wave one $S_{fi}^{(pw)}$ with $\psi_{i} ({\bm p}_{i}), \psi_f ({\bm p}_f)$ being the in/out wave functions. 
The number of events depends on $S_{fi}^{(pw)} ({\bm p}){S_{fi}^*}^{(pw)} ({\bm p}^{\prime})$ integrated over all pairs of momenta, ${\bm p}$ and ${\bm p}^{\prime}$,
with some weights. 
Such products of amplitudes cannot, as a rule, 
be reduced to the standard traces and the terms such as a pair of spinors $u({\bm p}) \bar{u}({\bm p}^{\prime})$ have to be expanded over a complete set of $16$ Dirac matrices. 
This makes the customary calculation procedure rather cumbersome and technically challenging (although not impossible per se).

Conversely, both the momenta coincide in the plane-wave limit when $S_{fi}^{(pw)} ({\bm p}) {S_{fi}^*}^{(pw)} ({\bm p}^{\prime}) \rightarrow |S_{fi}^{(pw)} ({\bm p})|^2$. 
Therefore when the packets are narrow, $\sigma_p \ll \langle p \rangle$, the following inequality holds true:
\begin{eqnarray}
& \displaystyle
\frac{|{\bm p} + {\bm p}^{\prime}|}{2} \gg |{\bm p} - {\bm p}^{\prime}|.
\label{inEq}
\end{eqnarray}
A density matrix written in these variables, $({\bm p} + {\bm p}^{\prime})/2$ and ${\bm p} - {\bm p}^{\prime}$, is called the Wigner function, introduced by Wigner in $1932$ \cite{Wigner}.

It is this inequality that allows one to develop a perturbation theory in which the ratio $\sigma_p/\langle p\rangle$ 
serves as a small parameter, the in/out-states are described by their Wigner functions, and the number of events, a luminosity, 
and the cross section are expanded into $\sigma_p/\langle p\rangle$-series. When $\sigma_p \rightarrow 0$, 
the particle's wave front turns flat and the phase of its wave function does not contribute to the observables.
Conversely, for any finite $\sigma_p$ the cross section gets corrections that depend explicitly on phases of the incoming states. 

\subsection{What other non-plane-wave effects can one expect?}

Naively one would think that non-plane-wave corrections to the (plane-wave) cross section are attenuated as $\lambda_c^2/\sigma_x^2 \ll 1$ 
where $\lambda_c = \hbar/(mc)$ is a Compton wave length of a particle and $\sigma_x \sim 1/\sigma_p$. 
As we shall demonstrate, at least some of them indeed are.
There are, however, corrections with a more complex model-dependent behavior. For instance, for any finite $\sigma_p$ 
the observables grow dependent not only on the absolute value of the amplitude $|M_{fi}^{(pw)}|$ but also on its phase $\zeta_{fi}$.
The problem of determining the phase of scattering amplitudes is of high importance for hadronic physics and has a long history (see, for example, \cite{West}). 
The ratio $\rho (s, t) = \Re\, M_{fi}^{(pw)}/\Im\, M_{fi}^{(pw)} = \cot \zeta_{fi} (s,t)$ is calculated in different models, including Regge approaches, 
and it is extracted from elastic $pp-$ and $p\bar{p}$-collisions. It is known that at the small transferred momenta $|t|$ 
this function is small, $|\rho (s, t)| \ll 1$, on the energy scale from several GeV to several TeV \cite{Cudell, Sel, Sel_12, Dremin}, 
thus manifesting the high value of the phase $\zeta_{fi}$. A proper analysis of the elastic proton-proton collisions at the Large Hadron Collider (LHC) at $\sqrt{s} = 7$ TeV
by the TOTEM collaboration \cite{TOTEM} has also shown that the real part of the amplitude can dominate for the large transferred momenta \cite{Dremin_PRD}.
Finally, just a few months ago the TOTEM collaboration managed to estimate the amplitude's phase at $\sqrt{s} = 8$ TeV in the interval of $|t|$ from $6\times10^{-4}\, \text{GeV}^2$ to $0.2\, \text{GeV}^2$
with a much better accuracy than before \cite{TOTEM_2016}.

As we have recently shown \cite{EPL} and demonstrate in more detail in this paper, 
scattering beyond the plane-wave approximation does not allow one to extract the Coulomb- or hadronic phase itself, 
but it does allow us to estimate how these phases change with $t$ or with the scattering angle $\theta_{sc}$.
We make an analysis that does not depend on phases of the incoming states.
First steps towards this direction have been taken by Ivanov for vortex beams in \cite{I_Phase}.

The paper is organized as follows. In Sec.$2$ we give general formulas for the probability and the cross section within the Wigner formalism.
We calculate several different Wigner functions in Sec.$3$, including the ones of a vortex particle and of an Airy state. 
In the latter case we compare an exact Wigner function with the corresponding approximate expression derived when $\sigma_p/\langle p \rangle \ll 1$. 
In Sec.$3.7$ we give a general approximate formula for such a function of a wave packet with an arbitrary complex phase.
The Lorentz invariant generalizations are given in Sec.$3.8$.
Sec.$4$ is devoted to derivation of a general probability formula when the in-states represent the wave packets with phases.
The corresponding generalization for beams is presented in Sec.$4.5$. The first non-vanishing corrections to the plane-wave results
are given in Secs.$4.6, 4.7$. A QED example is given in Sec.$4.8$. 
The effects of the scattering amplitude's phase are discussed in Sec.$5$. 
In Sec.$6$ we solve a similar problem but when the out-states are described as the Bessel ones with some OAM.
We summarize in Sec.$7$.

In what follows the terms ``a wave function'' and ``a phase'' will refer mostly to the momentum representation, and we mark $\sigma \equiv \sigma_p$ everywhere. 
As we do not put any limitations on the scattering amplitude, we shall use the term ``scattering'', for the sake of conciseness, 
for all the elastic and inelastic $2 \rightarrow N_f$ processes.
The units $\hbar = c = 1$ are used.


\section{Probability and cross section}\label{Gen.}

Let us consider a generic (not necessarily elastic) scattering of two wave packets with $N_f$ final plane waves with the momenta ${\bm p}_3, {\bm p}_4, ..., {\bm p}_{N_f+2}$. 
As demonstrated by Kotkin \textit{et al.} \cite{Impact}, the scattering probability can be represented as a functional of the (generalized) cross section $d \sigma ({\bm k}, {\bm p}_{1,2})$
and a function that we shall denote $\mathcal L ({\bm k}, {\bm p}_{1,2})$ and call \textit{the particle correlator}:
\begin{eqnarray}
& \displaystyle
dW = |S_{fi}|^2\, \prod\limits_{f=3}^{N_f+2} V \frac{d^3 p_f}{(2\pi)^3} =  \int \frac{d^3 p_1}{(2\pi)^3}\frac{d^3 p_2}{(2\pi)^3}\frac{d^3 k}{(2\pi)^3}\,\, d \sigma ({\bm k}, {\bm p}_{1,2})\, \mathcal L^{(2)} ({\bm k}, {\bm p}_{1,2}),
\cr & \displaystyle
d \sigma ({\bm k}, {\bm p}_{1,2}) = (2\pi)^4\, \delta \Big (\varepsilon_1 ({\bm p}_1 + {\bm k}/2) + \varepsilon_2 ({\bm p}_2 - {\bm k}/2) - \varepsilon_f \Big )\, \delta^{(3)} ({\bm p}_1 + {\bm p}_2 - {\bm p}_f) 
\cr & \displaystyle 
\times T_{fi}^{(pw)} ({\bm p}_1 + {\bm k}/2, {\bm p}_2 - {\bm k}/2) {T_{fi}^*}^{(pw)} ({\bm p}_1 - {\bm k}/2, {\bm p}_2 + {\bm k}/2) \frac{1}{\upsilon ({\bm p}_1, {\bm p}_2)}\prod\limits_{f=3}^{N_f+2} \frac{d^3 p_f}{(2\pi)^3},
\cr & \displaystyle 
\mathcal L ({\bm k}, {\bm p}_{1,2}) = \upsilon ({\bm p}_1, {\bm p}_2) \int\, dt\, d^3 r\,d^3 R\, e^{i{\bm k}{\bm R}}\, n_1 ({\bm r}, {\bm p}_1, t) n_2 ({\bm r} + {\bm R}, {\bm p}_2, t),
\label{M}
\end{eqnarray}
where 
\begin{eqnarray}
& \displaystyle 
\upsilon ({\bm p}_1, {\bm p}_2) = \frac{\sqrt {(p_1 p_2)^2 - m_1^2 m_2^2}}{\varepsilon_1 ({\bm p}_1) \varepsilon_2 ({\bm p}_2)} = \sqrt{({\bm u}_1 - {\bm u}_2)^2 - [{\bm u}_1 \times {\bm u}_2]^2},
\cr & \displaystyle 
\varepsilon_f = \sum\limits_{i=3}^{N_f+2}\varepsilon_i ({\bm p}_i),\,\, \varepsilon ({\bm p}) = \sqrt{{\bm p}^2 + m^2},\
{\bm p}_f = \sum\limits_{i=3}^{N_f+2}{\bm p}_i,\,\, {\bm u}_{1,2} = {\bm p}_{1,2}/\varepsilon_{1,2} ({\bm p}_{1,2}),
\label{Ma}
\end{eqnarray}
the amplitudes 
$$
T_{fi}^{(pw)} = \frac{M_{fi}^{(pw)}}{\sqrt{2 \varepsilon_1 2 \varepsilon_2 \prod\limits_{f=3}^{N_f+2}2\varepsilon_f}}
$$
no longer depend on the normalization volume $V$, and $n ({\bm r}, {\bm p}, t)$ is a (bosonic part of a) particle's Wigner function with the following properties:
\begin{eqnarray}
& \displaystyle
\int d^3r\, n ({\bm r}, {\bm p}, t) = |\psi ({\bm p}, t)|^2,\ \int \frac{d^3p}{(2\pi)^3}\, n ({\bm r}, {\bm p}, t) = |\psi ({\bm r}, t)|^2,\
\int \frac{d^3p}{(2\pi)^3}\, d^3 r\, n ({\bm r}, {\bm p}, t) = 1.
\label{W}
\end{eqnarray}
Note that in this approach we do not need fermionic Wigner functions 
(see, for example, \cite{BB_2, BB_3}), because the spin parts of all wave functions are factorized and enter into the amplitude $T_{fi}^{(pw)}$.
That is why the Wigner functions that we shall use in this paper are Lorentz scalars.


The function $d \sigma ({\bm 0}, {\bm p}_{1,2})$ coincides with a conventional definition of the plane-wave cross section. 
The dependence on ${\bm k}$ appears because of translational non-invariance
of the Wigner function and in the plane-wave case this invariance is recovered (see below).
For the well-normalized Wigner functions, the probability (\ref{M}) represents an unambiguous quantity in a sense that it does not depend on the auxiliary normalization variables, 
such as $V$ and $T$. When making a comparison with the plane-wave case, it is convenient, however, to define also \textit{the effective cross section} by dividing the probability by a luminosity factor $L$,
\begin{eqnarray}
& \displaystyle d\sigma = \frac{dW}{L},\ L = \int \frac{d^3 p_1}{(2\pi)^3}\frac{d^3 p_2}{(2\pi)^3}\frac{d^3 k}{(2\pi)^3}\,\, \mathcal L ({\bm k}, {\bm p}_{1,2}) = 
\cr & \displaystyle
= \int \frac{d^3 p_1}{(2\pi)^3}\frac{d^3 p_2}{(2\pi)^3}\,\, dt d^3 r\,\, \upsilon ({\bm p}_1, {\bm p}_2)\, n_1 ({\bm r}, {\bm p}_1, t) n_2 ({\bm r}, {\bm p}_2, t).
\label{Cr}
\end{eqnarray}
Note that the quantities $dW, L$, and $d\sigma$ are Lorentz-invariant, whereas the correlator is not.

Probability formula when all the particles, including the final ones, are not plane waves but rather some generic wave packets with the phases 
can be derived following the very same procedure as described in Ref.\cite{Impact}. For a special case with only two final states, the result reads:
\begin{eqnarray}
& \displaystyle
dW = |S_{fi}|^2\, dn_f = \int \prod\limits_{i=1}^{4}\frac{d^3 p_i}{(2\pi)^3}\prod\limits_{j=1}^{4}\frac{d^3 k_j}{(2\pi)^3}\,\, (2\pi)^3 \delta^{(3)} ({\bm k}_1 + {\bm k}_2 - {\bm k}_3 - {\bm k}_4)\,\, d \sigma \left ({\bm k}, {\bm p}\right )\, \mathcal L^{(4)} \left ({\bm k}, {\bm p}\right ),
\cr & \displaystyle
d \sigma \left ({\bm k}, {\bm p}\right ) = (2\pi)^4\, \delta \Big (\varepsilon_1 ({\bm p}_1 + {\bm k}_1/2) + \varepsilon_2 ({\bm p}_2 + {\bm k}_2/2) - \varepsilon_3 ({\bm p}_3 + {\bm k}_3/2) - \varepsilon_4 ({\bm p}_4 + {\bm k}_4/2) \Big ) 
\cr & \displaystyle 
\times\, \delta^{(3)} ({\bm p}_1 + {\bm p}_2 - {\bm p}_3 - {\bm p}_4)\, T_{fi}^{(pw)} \left ({\bm p} + {\bm k}/2 \right ) {T_{fi}^*}^{(pw)} \left ({\bm p} - {\bm k}/2 \right ) \frac{1}{\upsilon ({\bm p}_1, {\bm p}_2)}\, dn_f,
\cr & \displaystyle 
\mathcal L \left ({\bm k}, {\bm p}\right ) = \upsilon ({\bm p}_1, {\bm p}_2) \int\, dt\, d^3 r_1\, d^3 r_2\, d^3 r_3\, d^3 r_4\, \exp\left\{-i{\bm k}_2{\bm r}_2 + i{\bm k}_3{\bm r}_3 + i{\bm k}_4{\bm r}_4\right\} \times
\cr & \displaystyle 
n_1 ({\bm r}_1, {\bm p}_1, t) n_2 ({\bm r}_1 + {\bm r}_2, {\bm p}_2, t)\, n_3 ({\bm r}_1 + {\bm r}_3, {\bm p}_3, t),\, n_4 ({\bm r}_1 + {\bm r}_4, {\bm p}_4, t),
\label{M4}
\end{eqnarray}
where $\mathcal L$ represents a \textit{$4$-particle correlator}, $dn_f$ is an integration measure for the final states, ${\bm p}$ and ${\bm k}$ 
denote the sets of all vectors: ${\bm p}_1, {\bm p}_2, {\bm p}_3, {\bm p}_4$ and ${\bm k}_1, {\bm k}_2, {\bm k}_3, {\bm k}_4$, respectively. 
This expression can be useful when the final detected state is characterized not with the $3$-momenta and probably with spins, but with a different set of quantum numbers. 
In this case the measure $dn_f$ may also include discrete variables such as the OAM. A need for spatial- and temporal localization of the detected states appears, for instance, in the theory of neutrino oscillations \cite{Akhmedov_09, Akhmedov_10, Akhmedov_Found} or in QED calculations with the twisted photons \cite{Serbo}.

\section{Wigner functions}

\subsection{Generalities}

Before we turn to scattering, let us take a closer look at the Wigner functions and their properties. 
If the system is in a pure state with a wave function $\psi ({\bm p})$, then its Wigner function 
can be found as follows (see, for example, a good pedagogical introduction by Case \cite{Case} or the textbook \cite{deGr})
\begin{eqnarray}
& \displaystyle
n ({\bm r}, {\bm p}, t) = \int \frac{d^3k}{(2\pi)^3}\,\, e^{i{\bm k}{\bm r}}\, \psi^* ({\bm p} - {\bm k}/2, t) \psi ({\bm p} + {\bm k}/2, t),
\label{WA}
\end{eqnarray}
where 
$$
\psi ({\bm p}, t) = \psi ({\bm p})\, \exp\{-i t\, \varepsilon ({\bm p})\}.
$$
In what follows we shall derive and analyse several examples of such functions.
Since the integral in (\ref{WA}) cannot be always evaluated exactly, we shall develop a method for obtaining a suitable approximate expression,
applicable when $\sigma/\langle p \rangle \ll 1$. The literature on the Wigner functions of optical vortices, Airy beams and their generalizations is extensive --- 
see, for example, Refs.\cite{Alonso, Wigner_vortex, Wigner_Airy, Barnett_2015}. 
Here we do not intend to make a comprehensive comparison of our results with those obtained in optics.

In this paper we employ only wave packets with a Gaussian envelope in momentum representation, bearing in mind that a wide class of packets 
can actually be approximated by this form (see discussion in Ref.\cite{Akhmedov_10}).
Conversely, the corresponding wave functions in configuration space may appear to be non-Gaussian, because of the phases.
Parameters characterizing the packets are: a mean momentum $\langle {\bm p}\rangle$, a momentum uncertainty $\sigma$, and a phase $\varphi ({\bm p})$.
While the latter is Lorentz invariant, a dispersion $\sigma^2$ generally transforms under the Lorentz transformation not as a scalar, but as a $3$-vector:
\begin{eqnarray}
& \displaystyle
\sigma_i^2 \sim \left \langle \left (p_i - \langle p_i\rangle \right )^2 \right\rangle.
\label{sigmagen}
\end{eqnarray}
We also introduce a $3$-tensor 
$$
\sigma_{ij} = \diag \{\sigma_x, \sigma_y, \sigma_z\},
$$ 
which can be non-diagonal but still symmetric in an arbitrary frame of reference. 
For a boost along, say, z axis we have
\begin{eqnarray}
& \displaystyle
\sigma_{x,y} = \sigma_{x,y,}^{\prime},\ \sigma_z = \gamma\, \sigma_z^{\prime}\ \text{with}\ \gamma = \varepsilon ({\bm p})/m,
\label{sigmatr}
\end{eqnarray}
with $\sigma_i^{\prime}$ in a frame where the packet is at rest on average. It is much more illustrative, however, 
to start with the customary ``non-relativistic'' expressions for packets with the only one $\sigma$. 
That is to say, we first work in the frame of reference in which
\begin{eqnarray}
& \displaystyle
\sigma_x = \sigma_y = \sigma_z \equiv \sigma.
\label{sigmaframe}
\end{eqnarray}
Lorentz invariant generalizations for packets with $\sigma \rightarrow \sigma_{ij}$ are straightforward, and it will be done at the very last stage in Sec.\ref{Lor}.

\subsection{Wigner function of a Gaussian wave packet}

Let us consider a Gaussian wave packet with the following wave function:
\begin{eqnarray}
& \displaystyle
\psi ({\bm p}) = \pi^{3/4} \Big (\frac{2}{\sigma}\Big )^{3/2} \exp\left\{-i{\bm r}_0 {\bm p} - \frac{({\bm p} - \langle{\bm p}\rangle)^2}{2\sigma^2}\right\},\ \int \frac{d^3p}{(2\pi)^3}\, |\psi ({\bm p}, t)|^2 = 1
\label{psiA}
\end{eqnarray}
where ${\bm r}_0$ denotes initial conditions. Lorentz invariance of the normalization requires that $\sigma^3\, (\equiv \det \sigma)$ transform as an inverse volume:
\begin{eqnarray}
& \displaystyle
\gamma^{-1}\,\sigma^3 = \text{inv},
\label{sigmainv}
\end{eqnarray}
in accordance with Eq.(\ref{sigmatr}).

Substituting (\ref{psiA}) into Eq.(\ref{WA}),
we see that the main contribution to the integral comes from the small values of ${\bm k}$: $|{\bm k}| \lesssim \sigma$ (recall the Ineq.(\ref{inEq})):
\begin{eqnarray}
& \displaystyle
\psi^* ({\bm p} - {\bm k}/2, t) \psi ({\bm p} + {\bm k}/2, t) \propto \exp \left\{-\frac{{\bm k}^2}{(2\sigma)^2} - \frac{({\bm p} - \langle{\bm p}\rangle)^2}{\sigma^2}\right\}
\label{propto}
\end{eqnarray}
Therefore, we can make the following expansion
\begin{eqnarray}
& \displaystyle
\varepsilon ({\bm p} + {\bm k}/2) - \varepsilon ({\bm p} - {\bm k}/2) \approx \frac{{\bm  p} {\bm k}}{\varepsilon ({\bm p})} + \mathcal O (k^3)  = {\bm u} ({\bm p}) {\bm k} + \mathcal O (k^3)
\label{varepsilonA}
\end{eqnarray}
Within this accuracy, we arrive at the simple expression (compare with that for coherent states \cite{Caruthers, deGr, Zav})
\begin{eqnarray}
& \displaystyle
n ({\bm r}, {\bm p}, t) = 8\, \exp\left\{-\frac{({\bm p} - \langle{\bm p}\rangle)^2}{\sigma^2} - \sigma^2 ({\bm r} - \langle{\bm r} \rangle )^2\right\} 
\label{WGaussA}
\end{eqnarray}
where 
\begin{eqnarray}
& \displaystyle
\langle{\bm r} \rangle = {\bm r}_0 + {\bm u} ({\bm p}) t.
\label{r}
\end{eqnarray}
As a cautionary remark, we note that this is not yet the true mean path of the system, as it still depends on ${\bm p}$, not $ \langle{\bm p} \rangle$. 
Although the function (\ref{WGaussA}) itself does not spread, it is consistent with the momentum-coordinate uncertainty relations.
It also satisfies the continuity equation,
\begin{eqnarray}
& \displaystyle
\frac{\partial n ({\bm r}, {\bm p}, t)}{\partial t} = - {\bm u} \frac{\partial n ({\bm r}, {\bm p}, t)}{\partial {\bm r}},
\label{WEqA}
\end{eqnarray}
and represents a relativistic $(3+1)$ D generalization (that is why $n ({\bm r}, {\bm p}, t) \leq 2^3$) of the corresponding function of a quantum oscillator, 
and when $\hbar \rightarrow 0$ it describes the so-called quasi-classical trajectory-coherent state of a boson (see, for example, \cite{Bagrov_Mono, deGr, Bagrov_TCS}). 

The higher-order terms that we have neglected in (\ref{varepsilonA}), give corrections to (\ref{WGaussA}) of the order of $\sigma^4$. 
Indeed, taking into account the next term in the expansion (\ref{varepsilonA}),
\begin{eqnarray}
& \displaystyle
k_i k_j k_k \frac{1}{24 \varepsilon^2}\, (3 u_i u_j u_k - u_i \delta_{jk} - u_j \delta_{ik} - u_k \delta_{ij}),
\label{O3}
\end{eqnarray}
we obtain the corresponding correction to Eq.(\ref{WGaussA}),
\begin{eqnarray}
& \displaystyle
n ({\bm r}, {\bm p}, t) = 8\, \exp\left\{-\frac{({\bm p} - \langle{\bm p}\rangle)^2}{\sigma^2} - \sigma^2 ({\bm r} - \langle{\bm r} \rangle )^2\right\} 
\cr & \displaystyle 
\times \Big (1 + \sigma^4 \frac{t}{2\varepsilon^2} ({\bm u}, {\bm r} - \langle {\bm r}\rangle ) (3 {\bm u}^2 - 5) + \mathcal O(\sigma^6)\Big ),
\label{WGaussAO3}
\end{eqnarray}
where the second term in parentheses is supposed to be small compared to unity, 
and that is why this function stays everywhere positive in all the orders in $\sigma$.

In the plane-wave limit with $\sigma \rightarrow 0$, we find 
\begin{eqnarray}
& \displaystyle
n ({\bm r}, {\bm p}, t) \rightarrow \frac{\sigma^3}{\pi^{3/2}}\, (2\pi)^3 \delta \left({\bm p} - \langle{\bm p}\rangle\right) = \frac{\sigma^3}{\pi^{3/2}}\, |\psi ({\bm p})|^2,
\label{WPWA}
\end{eqnarray}
from where we get the following useful rule (recall Eq.(\ref{sigmainv}))
\begin{eqnarray}
& \displaystyle
\frac{\sigma^3}{\pi^{3/2}} \rightarrow \frac{1}{V} = {j^0}^{(pw)}.
\label{Vrule}
\end{eqnarray}

\subsection{Wigner function of a Gaussian beam}\label{Wbeam}

Having obtained the everywhere positive Wigner function of a Gaussian wave packet, we can now derive a similar expression for a beam of $N_b$ identical non-interacting particles.
Such a function is a result of the statistical averaging of the wave-packets with some distribution over the packets' centers, $f({\bm r}_0)$:
\begin{eqnarray}
& \displaystyle
n_b ({\bm r}, {\bm p}, t) = N_b \int d^3 {\bm r}_0\, n ({\bm r}, {\bm p}, t; {\bm r}_0) f({\bm r}_0),\ \int d^3{\bm r}\frac{d^3{\bm p}}{(2\pi)^3}\, n_b ({\bm r}, {\bm p}, t) = N_b,
\label{nbeam}
\end{eqnarray}
and we imply that the distance between the particles, $\sim \sigma_b/N_b$, does not exceed the coherence length of one wave-packet $1/\sigma$,
that is,
\begin{eqnarray}
& \displaystyle
N_b \gtrsim \text{or} \gg \sigma\sigma_b.
\label{beamineq}
\end{eqnarray}
For a Gaussian beam with
\begin{eqnarray}
& \displaystyle
f({\bm r}_0) = \frac{1}{\pi^{3/2} \sigma_b^3} \exp\Big\{-\frac{({\bm r}_0 - {\bm r}_b)^2}{\sigma_b^2}\Big\}
\label{fG}
\end{eqnarray}
we find:
\begin{eqnarray}
& \displaystyle
n_b ({\bm r}, {\bm p}, t) = N_b\,\frac{8}{(1 + \sigma^2 \sigma_b^2)^{3/2}}\, \exp\Big\{-\frac{({\bm p} - \langle{\bm p}\rangle)^2}{\sigma^2} - \Sigma^2 \left ({\bm r} - {\bm r}_b - {\bm u} ({\bm p}) t \right )^2\Big\} 
\label{nbeamG}
\end{eqnarray}
where
\begin{eqnarray}
& \displaystyle
\Sigma^2 = \frac{\sigma^2}{1 + \sigma^2 \sigma_b^2} \equiv \mathcal O(\sigma_b^{-2})
\label{Sigma}
\end{eqnarray}
Note that for a beam, the momentum uncertainty $\sigma$ and the spatial width $\sigma_b$ are two independent parameters, 
and in the overwhelming majority of practical cases 
$$
\sigma_b \gg 1/\sigma
$$
Say, for the LHC proton beam with $\sigma_b \sim 10 \mu$m and $\sigma/\langle p \rangle \lesssim 1\%$ \cite{LHC}, 
we have $\sigma \sigma_b > 10^{11}$ and, therefore, $\Sigma^2 \approx 1/\sigma_b^2$. The case with $\sigma \sigma_b \sim 1$ is realized, 
for instance, for 300-keV electrons focused in a spot of 1 $\text{\AA}$ \cite{Angstrom}; and now $\Sigma^2 \approx 1/(2\sigma_b^2)$.


\subsection{Wigner function of a Bessel state}

A Bessel state is characterized with the following quantum numbers: 
the longitudinal momentum $p_{\parallel}$, an absolute value of the transverse momentum $\kappa$, the energy $\varepsilon (\kappa, p_{\parallel})$, 
and a projection of the OAM onto the propagation axis, $L_z \equiv \ell$ (see, for example, \cite{Bliokh_07, Bliokh_11, I_PRD, PRA_12}). 
The wave function is
\begin{eqnarray}
& \displaystyle
\psi ({\bm p}) = (2 \pi)^{3/2} \sqrt{\frac{\pi}{R L}}\, \frac{\delta (p_{\perp} - \kappa)}{\sqrt{p_{\perp}}}\, \delta (p_z - p_{\parallel}) e^{i\ell\, \phi_p},\ \int \frac{d^3p}{(2\pi)^3}\, |\psi ({\bm p})|^2 = 1,
\label{psiB}
\end{eqnarray}
where $\phi_p$ is the azimuthal angle, and one can use the following rules
\begin{eqnarray}
& \displaystyle
(\delta (p_{\perp} - \kappa))^2 \rightarrow \frac{R}{\pi}\, \delta (p_{\perp} - \kappa),\ (\delta (p_z - p_{\parallel}))^2 \rightarrow \frac{L}{2\pi}\,\delta (p_z - p_{\parallel}).\label{rulesB}
\end{eqnarray}
The corresponding Wigner function can be found exactly applying Eq.(\ref{WA}):
\begin{eqnarray}
& \displaystyle
n ({\bm r}, {\bm p}, t; \ell) \equiv n ({\bm r}, {\bm p}; \ell) = \frac{4\pi}{RL} \frac{\Theta (\kappa - p_{\perp})}{p_{\perp} \sin \xi}\, \delta (p_z - p_{\parallel})\,
\cos {\Big (2 \ell \xi - [{\bm r} \times {\bm p}]_z\, 2\tan \xi \Big )},
\label{WBess}
\end{eqnarray}
where
\begin{eqnarray}
& \displaystyle
\sin \xi = \sqrt{1 - (p_{\perp}/\kappa)^2},\, \cos \xi = p_{\perp}/\kappa,\, \tan \xi = \sqrt{(\kappa/p_{\perp})^2 - 1},
\label{xi}
\end{eqnarray}
and the Heaviside function is extended so that $\Theta (0) = 1$. 
This Wigner function can be negative, it is invariant under the Lorentz boosts along the z axis, 
and it coincides up to a common factor with the so-called Wolf function in optics (see Eq.(22) in Ref.\cite{Barnett_2015}).

As can be easily shown,
\begin{eqnarray}
& \displaystyle
\int \frac{d^3 p}{(2\pi)^3}\, n ({\bm r}, {\bm p}, t; \ell) = \text{const}\, J_{\ell}^2 (\kappa \rho),
\label{dens}
\end{eqnarray}
as should be for an azimuthally symmetric Bessel state. Hence, the maximum value of $\ell$ is
\begin{eqnarray}
& \displaystyle
\ell_{\text{max}} \sim \kappa \rho.
\label{lBessmax}
\end{eqnarray} 

The singularity $\xi \rightarrow 0$ is integrable, and all the properties of a generic Wigner function (\ref{W}) hold true. The lack of time dependence means the absence of spreading, 
the well-known feature of the Bessel beams. Unlike the wave function (\ref{psiB}), the Wigner function (\ref{WBess}) can possess any transverse momentum up to $\kappa$, 
and it is just maximized when $p_{\perp}$ reaches $\kappa$.

\subsection{Wigner function of a packet with OAM}

If an OAM eigenstate has a distribution over the transverse momentum,
\begin{eqnarray}
& \displaystyle
\psi ({\bm p}) = (2 \pi)^{3/2} \sqrt{\frac{2}{L \sigma^2}}\, \delta (p_z - p_{\parallel}) \exp \Big \{-\frac{({\bm p}_{\perp} - {\bm \kappa})^2}{2\sigma^2} - i{\bm r}_{0,\perp}{\bm p}_{\perp} + i\ell \phi\Big\},
\label{psiC}
\end{eqnarray}
then such a state, unlike the pure Bessel one, has a finite OAM dispersion (or the OAM spectrum) when ${\bm \kappa} \ne 0$ -- see, for example, Refs.\cite{Torres_2003, Barnett, PRA}.
When calculating the corresponding Wigner function with the use of Eq.(\ref{WA}), we need to deal with the factor
$$
\exp\{i\ell (\phi_+ - \phi_-)\},
$$
where $\phi_{\pm}$ are the azimuthal angles of the vectors ${\bm p} \pm {\bm k}/2$. By analogy with Eq.(\ref{varepsilonA}), we make an expansion of this over the small ${\bm k}$:
\begin{eqnarray}
& \displaystyle
\phi_+ - \phi_- \approx {\bm k}\, \frac{\hat{{\bm z}} \times {\bm p}}{{\bm p}_{\perp}^2} + O (k^3)
\label{expansionphi}
\end{eqnarray}
This leads to the following result
\begin{eqnarray}
& \displaystyle
n ({\bm r}, {\bm p}, t; \ell) = 8\, \frac{\pi}{L}\,\delta (p_z - p_{\parallel})\, \exp\left\{-\frac{({\bm p}_{\perp} - {\bm \kappa})^2}{\sigma^2} - \sigma^2 \left ({\bm r}_{\perp} - \langle{\bm r} \rangle_{\perp} + \ell\, \frac{\hat{\bm z}\times{\bm p}}{{\bm p}_{\perp}^2} \right )^2\right\}
\label{WWP}
\end{eqnarray}
with $\langle{\bm r} \rangle_{\perp}$ being a transverse part of (\ref{r}). 
When the longitudinal momentum has also the same uncertainty, 
we arrive at the generalization of the coherent state (\ref{WGaussA}):
\begin{eqnarray}
& \displaystyle
n ({\bm r}, {\bm p}, t; \ell) = 8\, \exp\left\{-\frac{({\bm p} - \langle{\bm p}\rangle)^2}{\sigma^2} - \sigma^2 \left ({\bm r} - \langle{\bm r} \rangle + \ell\, \frac{\hat{\bm z} \times {\bm p}}{{\bm p}_{\perp}^2} \right )^2\right\}
\label{WGaussOAM}
\end{eqnarray}
As can be seen, the wave packet with the OAM implies, similar to the pure Bessel state, a finite transverse momentum
and in the plane-wave limit, $\sigma \rightarrow 0$, the OAM vanishes. We would like to emphasize that even for a state with $\langle{\bm p}\rangle_{\perp} = 0$, 
the mean absolute value of the transverse momentum appears to be non-vanishing, 
$$
\langle p_{\perp}\rangle \sim \sigma,
$$
as can be readily checked (see Eq.(40) in \cite{Born}). That is why it is sometimes helpful to think of $\sigma$ as of the transverse momentum. 
The maximum value of the OAM is
\begin{eqnarray}
& \displaystyle
\ell_{\text{max}} \sim p_{\perp}/\sigma \sim 1, \ \text{or for a beam:}\ \ell_{\text{max}} \sim p_{\perp}\sigma_b \sim \sigma\sigma_b.
\label{lpackmax}
\end{eqnarray}

\begin{figure}
\center
\includegraphics[width=16.00cm, height=6.50cm]{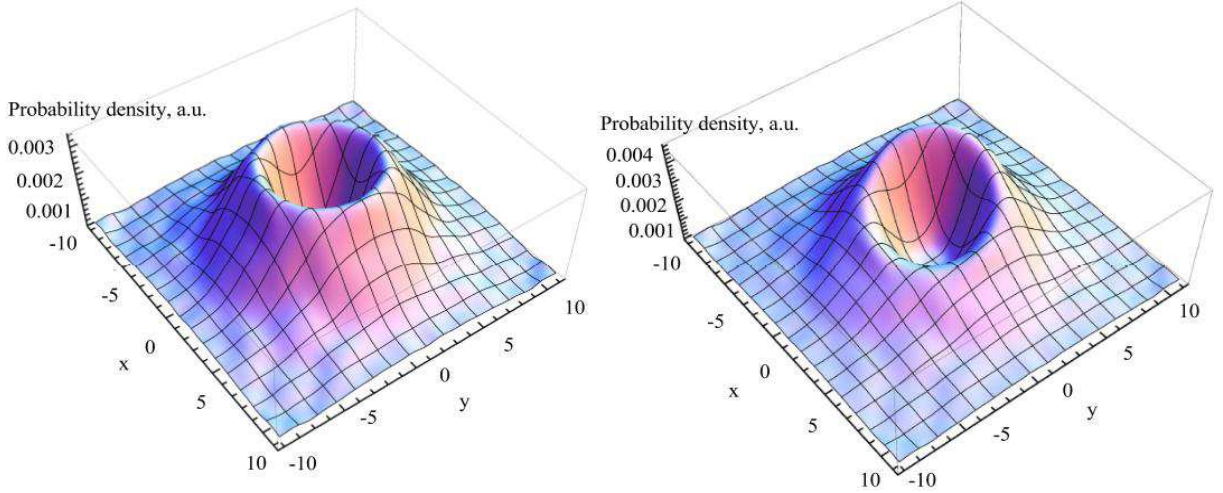}
\caption{Spatial distribution of the wave packet with OAM and the Wigner function from Eq.(\ref{WGaussOAM}). Parameters: $m = 1, \sigma/\langle p_z\rangle = 1/5,\, \ell = 5,\, {\bm r}_0 = z = t = 0$. \textit{Left panel}: $\langle {\bm p}\rangle_{\perp} = 0$. \textit{Right panel}: $\langle {\bm p}\rangle_{\perp} = \{0.1, 0.1\} \sigma$.
\label{Fig1}}
\end{figure}

Such a state with $\langle{\bm p}\rangle_{\perp} = 0$ has an azimuthally symmetric distribution of the intensity with a central minimum (see the left panel in Fig.\ref{Fig1}),
thus representing a well-normalized \textit{generalization of the pure Bessel state}. Unlike the Wigner function of the latter, however, Eq.(\ref{WGaussOAM}) is positive for all the values of $\ell$. 
The right panel in Fig.\ref{Fig1} describes a similar wave packet with a non-vanishing transverse momentum.
Such states can be useful for quantum entanglement in the OAM \cite{Mair_2001} and for probing the phase of the scattering amplitude 
in a collision experiment with vortex particles (see Sec.\ref{asymm}).

Performing the Weyl transformation for $\hat{L}_z$ operator, as explained for instance in Refs.\cite{deGr, Case}, we find the OAM expectation value
calculated with the use of these functions:
\begin{eqnarray}
& \displaystyle
\langle \hat{L}_z\rangle = [{\bm r}_0 \times \langle {\bm p}\rangle]_z + \ell,
\label{Lz}
\end{eqnarray}
as should be. And of course the OAM-dependent term in (\ref{WGaussOAM}) does not change the mean trajectory $\langle\hat{{\bm r}}\rangle$
calculated with this Wigner function.

If the OAM is quantized not relative to the z-axis, but along a unit vector $\hat{\bm n}$, then one should make the following substitution in the Eqs.(\ref{WWP}),(\ref{WGaussOAM}):
\begin{eqnarray}
& \displaystyle
\ell\, \frac{\hat{\bm z} \times {\bm p}}{{\bm p}_{\perp}^2} \rightarrow \ell\, \frac{\hat{\bm n} \times {\bm p}}{[\hat{\bm n} \times {\bm p}]^2}.
\label{subs}
\end{eqnarray}
This allows one to use packets with the so-called orbital helicity for which ${\bm n} = \langle{\bm p}\rangle/\sqrt{\langle {\bm p}\rangle^2}$ \cite{I_S}.

\begin{figure}
\centering
\includegraphics[width=16.00cm, height=6.00cm]{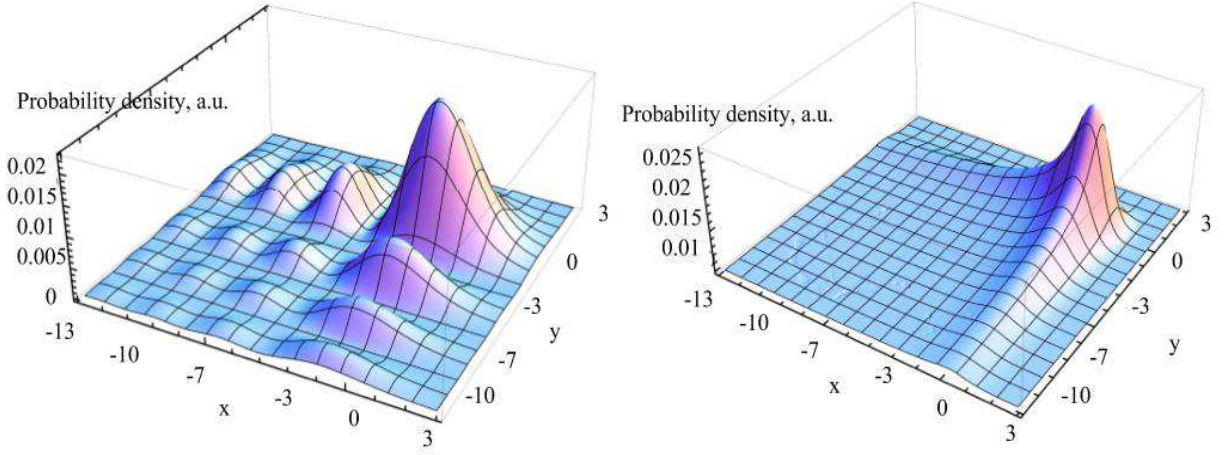}
\caption{Spatial distribution of the Airy wave packet. \textit{Left panel}: the one with the exact Wigner function from Eq.(\ref{WAiry}); \textit{Right panel}: the one with the approximate Wigner function from Eq.(\ref{WAiryapp}). 
Parameters: $m = 1, \sigma/\langle p\rangle_z = 1/5,\, \xi_x = \xi_y = 2/\sigma,\, {\bm r}_0 = z = t = \langle {\bm p}\rangle_{\perp} = 0$. 
\label{Fig2}}
\end{figure}

\subsection{Wigner function of an Airy particle}

The normalized wave function of an Airy particle is parameterized with a 2D vector\footnote{The widely used notation, ${\bm \xi} \rightarrow \{x_0, y_0\}$, 
is somewhat misleading because these parameters, $x_0$ and $y_0$, are not genuine initial conditions for the coordinates.} ${\bm \xi} = \{\xi_x, \xi_y\}$ \cite{Airy, Airy_beam}, 
which transforms as coordinates under the Lorentz boosts:
\begin{eqnarray}
& \displaystyle
\psi ({\bm p}) = \pi^{3/4} \Big ({\frac{2}{\sigma}}\Big )^{3/2}\, \exp \Big \{ - i{\bm r}_{0}{\bm p} -\frac{({\bm p} - \langle{\bm p}\rangle)^2}{2\sigma^2} + \frac{i}{3} \Big(\xi_x^3 p_x^3 + \xi_y^3 p_y^3\Big )\Big\}. \label{psiD}
\end{eqnarray}
The corresponding \textit{exact} Wigner function is found with the use of Eq.(\ref{WA}):
\begin{eqnarray}
& \displaystyle
n ({\bm r}, {\bm p}, t; {\bm \xi}) =  2^{13/3}\, \frac{\pi}{\sigma^2 \xi_x \xi_y} \exp \Big \{-\sigma^2 (z - \langle z\rangle)^2 -\frac{({\bm p} - \langle{\bm p}\rangle)^2}{\sigma^2} + 
\cr & \displaystyle 
+ \frac{1}{\sigma^2 \xi_x^3} \Big (x - \langle x\rangle + \xi_x^3 p_x^2 + \frac{1}{6\sigma^4 \xi_x^3}\Big ) + \frac{1}{\sigma^2 \xi_y^3} \Big (y - \langle y\rangle + \xi_y^3 p_y^2 + \frac{1}{6\sigma^4 \xi_y^3}\Big )\Big\},
\cr & \displaystyle
\times \text{Ai}\left[\frac{2^{2/3}}{\xi_x} \Big (x - \langle x\rangle + \xi_x^3 p_x^2 + \frac{1}{4 \sigma^4 \xi_x^3}\Big )\right ]
\text{Ai}\left[\frac{2^{2/3}}{\xi_y} \Big (y - \langle y\rangle + \xi_y^3 p_y^2 + \frac{1}{4 \sigma^4 \xi_y^3}\Big )\right ].
\label{WAiry}
\end{eqnarray}
This function may become negative together with the arguments of the Airy functions. 

It is instructive to obtain the corresponding approximate expression by employing the small-${\bm k}$ expansion,
\begin{eqnarray}
& \displaystyle
\frac{1}{3} \Big(\xi_x^3 (p_x + k_x/2)^3 + \xi_y^3 (p_y + k_y/2)^3\Big ) - \frac{1}{3} \Big(\xi_x^3 (p_x - k_x/2)^3 + \xi_y^3 (p_y - k_y/2)^3\Big ) \approx 
\cr & \displaystyle
\approx {\bm k} {\bm \eta} + \mathcal O(k^3),\quad
 {\bm \eta} \equiv {\bm \eta} ({\bm p}_{\perp}) = \{\xi_x^3 p_x^2, \xi_y^3 p_y^2, 0\}.
\label{phase}
\end{eqnarray}
As a result, we arrive at a simple everywhere-positive function:
\begin{eqnarray}
& \displaystyle
n ({\bm r}, {\bm p}, t; {\bm \xi}) = 8\, \exp\left\{-\frac{({\bm p} - \langle{\bm p}\rangle)^2}{\sigma^2} - \sigma^2 \left ({\bm r} - \langle{\bm r} \rangle + {\bm \eta} \right )^2\right\}.
\label{WAiryapp}
\end{eqnarray}
This demonstrates explicitly that the negative values of the Wigner function are connected with the non-Gaussian $\mathcal O(\sigma^4)$-terms
that we have neglected. An important distinction between the Airy phase and that of the vortex particle is that the fourth derivative of the former vanishes.
That is why the neglected $\mathcal O(\sigma^4)$-term is the only correction to this Wigner function. 

From Eq.(\ref{WAiryapp}) we infer:
\begin{eqnarray}
& \displaystyle
\xi_{\text{max}} \sim 1/\sigma,\ \text{or for a beam:}\ \xi_{\text{max}} \sim \frac{(1 + \sigma^2 \sigma_b^2)^{1/6}}{\sigma}.
\label{xiineq}
\end{eqnarray}

\begin{figure}
\centering
\includegraphics[width=16.00cm, height=6.00cm]{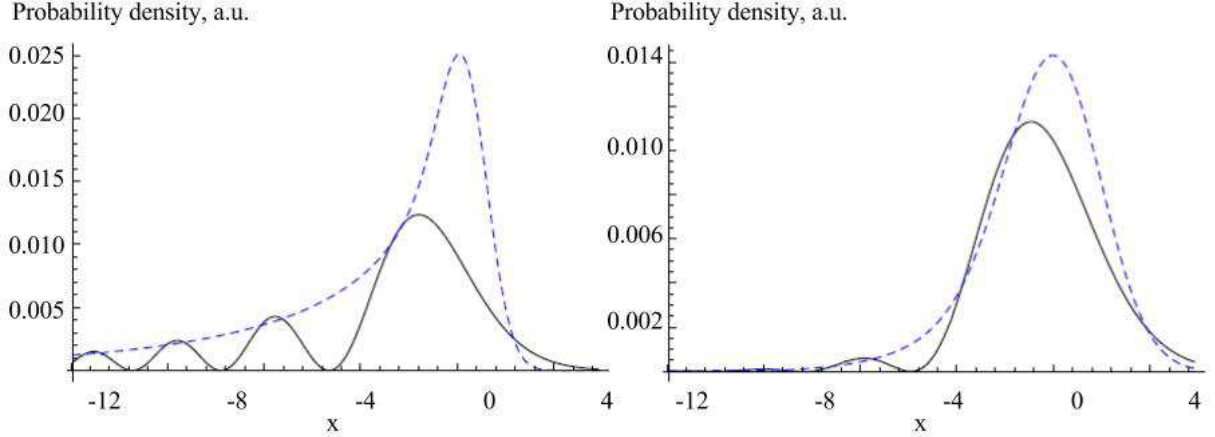}
\caption{Spatial distribution of the Airy wave packet: with the exact Wigner function (\ref{WAiry}) (black curve) vs. with the approximate one (\ref{WAiryapp}) (blue dashed curve). 
Parameters: $m = 1,\, \sigma/\langle p\rangle_z = 1/5,\, {\bm r}_0 = y = z = t = \langle {\bm p}\rangle_{\perp} = 0$. \textit{Left panel}: $\xi_x = \xi_y = 2/\sigma$, 
\textit{Right panel}: $\xi_x = \xi_y = 1/\sigma$. \label{Fig3}}
\end{figure}

In Figs.\ref{Fig2},\ref{Fig3} spatial distributions of the Airy wave packet are depicted for both the expressions: the exact- and approximate one.
As can be seen, a transit from the not-everywhere-positive Wigner function (\ref{WAiry}) 
to the everywhere-positive one (\ref{WAiryapp}) implies smoothing out the fast Airy oscillations for negative values of $x, y$,
and the approximate Wigner function works better when $\xi \lesssim \xi_{\text{max}} \sim 1/\sigma$.
Since the total areas under the both surfaces coincide, there are also the regions where the black curve in Fig.\ref{Fig3} may exceed the blue one (namely, when $x \ne 0, y\ne 0$). 

In collision of an Airy particle with another wave packet, these fast oscillations in Fig.\ref{Fig3} may play role when the latter is focused in a spot comparable to the oscillation period. 
Study of these effects obviously lies beyond the current experimental possibilities. In other words, although the differences in Figs.\ref{Fig2},\ref{Fig3} can be seen with the naked eye,  
the function (\ref{WAiryapp}) nevertheless represents \textit{a very good approximation} in the scattering problems.

\subsection{Wigner function of a packet with an arbitrary phase}

Comparing Eq.(\ref{WAiryapp}) with the one for the vortex state (\ref{WGaussOAM}), one can notice that they can be easily generalized 
for the wave function with an arbitrary, but regular complex phase $\varphi ({\bm p})$,
\begin{eqnarray}
& \displaystyle
\psi ({\bm p}) = \pi^{3/4} \Big ({\frac{2}{\sigma}}\Big )^{3/2}\, \exp \Big \{ - i{\bm r}_{0}{\bm p} -\frac{({\bm p} - \langle{\bm p}\rangle)^2}{2\sigma^2} + i \varphi ({\bm p})\Big\}. 
\label{psivarphi}
\end{eqnarray}
Then employing the same expansion,
\begin{eqnarray}
& \displaystyle
\varphi ({\bm p} + {\bm k}/2) - \varphi ({\bm p} - {\bm k}/2) \approx {\bm k}\, \frac{\partial \varphi ({\bm p})}{\partial{\bm p}} + \mathcal O(k^3),
\label{expphi}
\end{eqnarray}
we find the following Wigner function:
\begin{eqnarray}
& \displaystyle
n ({\bm r}, {\bm p}, t) = 8\, \exp\left\{-\frac{({\bm p} - \langle{\bm p}\rangle)^2}{\sigma^2} - \sigma^2 \left ({\bm r} - {\bm r}_0 - {\bm u} ({\bm p}) t + 
\frac{\partial \varphi ({\bm p})}{\partial{\bm p}}\right )^2\right\}.
\label{Wapp}
\end{eqnarray}
The similar expression for a Gaussian beam reads as:
\begin{eqnarray}
& \displaystyle
n_b ({\bm r}, {\bm p}, t) = N_b\,\frac{8}{(1 + \sigma^2 \sigma_b^2)^{3/2}}\, \exp\Big\{-\frac{({\bm p} - \langle{\bm p}\rangle)^2}{\sigma^2} - \Sigma^2 \left ({\bm r} - 
{\bm r}_b - {\bm u} ({\bm p}) t + \frac{\partial \varphi ({\bm p})}{\partial{\bm p}} \right )^2\Big\} 
\label{nbeamGPh}
\end{eqnarray}
with the vector ${\bm r}_b$ pointing to the center of the beam at $t=0$.

It is instructive to write down explicitly the condition of smallness of the higher-order terms that we neglected in Eq.(\ref{expphi}):
\begin{eqnarray}
& \displaystyle
\left |\frac{\partial \varphi}{\partial{\bm p}_i}\right | \gg \frac{\sigma^2}{6} \left |\frac{\partial^3 \varphi}{\partial {\bm p}_i \partial {\bm p}_j \partial {\bm p}_j}\right |,
\label{varphiineq}
\end{eqnarray}
For Airy beams, this yields just $\sigma^2 \ll {\bm p}^2 \sim \langle {\bm p}\rangle^2$. 
This inequality is obviously violated for the vortex particle 
with ${\bm p}_{\perp} = 0$ being a singularity in Eq.(\ref{expansionphi}). Consequently the expansion (\ref{expphi}) is not applicable 
when ${\bm p}_{\perp} \rightarrow 0$ or rather ${\bm p}_{\perp}^2 \ll \sigma^2$.
However the Wigner function (\ref{WGaussOAM}) itself is exponentially suppressed in this case as $\exp\{-\ell^2\sigma^2/{\bm p}_{\perp}^2\}$,
and that is why the formulas (\ref{WGaussOAM}), (\ref{Wapp}) can still be used in the entire ${\bm p}$-domain.

As before, the $\mathcal O(\sigma^4)$-corrections to these functions are responsible for the possible negativity. 
However in the paraxial regime with $\sigma \ll \langle p \rangle$, they are small, and a series like (\ref{WGaussAO3}) could never make the Wigner function not-everywhere-positive.
To put it simply, in scattering problems the possible negativity of the Wigner functions can reveal itself beyond the perturbative regime, 
that is, when $\sigma \sim \langle p\rangle$. This dictates focusing of the beam to a spot of 
$$
\sigma_b \sim 1/\sigma \sim 1/\langle p\rangle,
$$
which seems to be feasible only for cold systems with $\langle p\rangle \ll m$ or $\sigma_b \gg \lambda_c$ where $\lambda_c = 1/m \equiv \hbar/(mc)$ is a Compton wave length of a particle. 
In the overwhelming majority of practical cases, therefore, the everywhere-positive Wigner functions can be used with a good accuracy.

\subsection{Lorentz invariant generalizations}\label{Lor}

Lorentz invariance of the Wigner functions can be restored by making the following substitution in the wave packet (\ref{psiA}):
\begin{eqnarray}
\frac{1}{\sigma^{3/2}} \exp\left\{-\frac{({\bm p} - \langle {\bm p} \rangle )^2}{2\sigma^2}\right \} \rightarrow \frac{1}{\sqrt{\sigma_x \sigma_y \sigma_z}} \exp\left\{-\frac{(p_x - \langle p_x \rangle )^2}{2\sigma_x^2}-\frac{(p_y - \langle p_y \rangle )^2}{2\sigma_y^2}-\frac{(p_z - \langle p_z \rangle )^2}{2\sigma_z^2}\right \}.
\label{subs}
\end{eqnarray}
Then we have instead of Eq.(\ref{psiA}):
\begin{eqnarray}
& \displaystyle
\psi ({\bm p}) = \frac{\pi^{3/4} 2^{3/2}}{\sqrt{\det \sigma}} \exp\left\{-i{\bm p} {\bm r}_0 -\frac{1}{2}\, \left ({\bm p} - \langle {\bm p}\rangle \right ) \sigma^{-2} \left ({\bm p} - \langle{\bm p}\rangle \right )\right \}.
\label{expinv}
\end{eqnarray}
Here and in what follows 
$$
{\bm a}B{\bm b} \equiv a_{i}B_{ij}b_{j}.
$$ 
The terms like  $i{\bm p} {\bm r}_0$ can be made explicitly invariant by substituting $-i{\bm p} {\bm r}_0 \rightarrow i (\varepsilon t_0 - {\bm p} {\bm r}_0) \equiv i (pr_0)$ with $t_0 = 0$ in our case. The corresponding generalization for coherent states (\ref{WGaussA}) and (\ref{Wapp}) is 
\begin{eqnarray}
& \displaystyle
n ({\bm r}, {\bm p}, t) = 8 \exp\Big\{- \left ({\bm p} - \langle {\bm p}\rangle \right ) \sigma^{-2} \left ({\bm p} - \langle {\bm p}\rangle \right ) - \cr
& \displaystyle
\qquad \qquad - \left ({\bm r} - {\bm r}_0 - {\bm u}t + \frac{\partial \varphi}{\partial {\bm p}}\right)\sigma^2\left({\bm r} - {\bm r}_0 - {\bm u}t + \frac{\partial \varphi}{\partial {\bm p}}\right)\Big \}.
\label{CSgen}
\end{eqnarray}
and for a beam characterized with a symmetric matrix ${\sigma_b}_{ij}$ we arrive at the following formula instead of Eqs.(\ref{nbeamG}),(\ref{nbeamGPh}):
\begin{eqnarray}
& \displaystyle
n_b ({\bm r}, {\bm p}, t) = N_b\,\frac{8}{\sqrt{\det (1 + \sigma_{b}^2 \sigma^2)}}\, \exp\Big\{- \left ({\bm p} - \langle {\bm p}\rangle \right ) \sigma^{-2} \left ({\bm p} - \langle {\bm p}\rangle \right ) - 
\cr
& \displaystyle
\qquad \qquad - \left ({\bm r} - {\bm r}_b - {\bm u}t + \frac{\partial \varphi}{\partial {\bm p}}\right)\Sigma^2\left ({\bm r} - {\bm r}_b - {\bm u}t + \frac{\partial \varphi}{\partial {\bm p}}\right)\Big\}
\label{nbeamGLor}
\end{eqnarray}
with
\begin{eqnarray}
& \displaystyle
\Sigma^{-2}_{ij} = \sigma^{-2}_{ij} + {\sigma_b}_{ij}^2,\quad \text{which is}\quad \Sigma^{-2}_{ij} = \diag \left\{\frac{1 + \sigma_x^2 \sigma_{b,x}^2}{\sigma_x^2}, \frac{1 + \sigma_y^2 \sigma_{b,y}^2}{\sigma_y^2}, \frac{1 + \sigma_z^2 \sigma_{b,z}^2}{\sigma_z^2}\right \}
\label{Sigmaij}
\end{eqnarray}
when the matrices $\sigma, \sigma_b$ are diagonal.

An analogous expression for an Airy particle can be readily guessed from Eq.(\ref{WAiry}) 
and we encourage the reader to make this generalization.

\section{Non-plane-wave scattering}

\subsection{Generalities}

Having studied properties of the Wigner functions, we intend now to substitute these formulas into the general expression for the number of events from Sec.\ref{Gen.}
and then to expand it into series with a small parameter of $\sigma/\langle p \rangle \ll 1$. As we shall see, the first term in this expansion represents the conventional plane-wave result,
and the corrections to it embrace effects of finite monochromaticity of the incoming beams and of their spreading with time, of a finite impact-parameter, of phases of the incoming states, 
as well as of the general phase of the scattering amplitude. The three latter effects appear thanks to finite \textit{overlap} of the incoming wave packets,
which is proportional to
\begin{eqnarray}
& \displaystyle
\alpha^{-1} = \left (\frac{1}{2\sigma_1^2} + \frac{1}{2\sigma_2^2}\right )^{-1} = \frac{2\sigma_1^2 \sigma_2^2}{\sigma_1^2 + \sigma_2^2}.
\label{alpha}
\end{eqnarray}
This constant serves as a small parameter for such interference phenomena.
As before, we start in the frame of reference in which $\sigma_{ij} = \sigma \delta_{ij}$. 
Generalizations for arbitrary frames will be made in Sec.\ref{LorSc}. 
In this case the overlap is described by the following matrix:
\begin{eqnarray}
& \displaystyle
\alpha^{-1} \rightarrow {\alpha}_{ij}^{-1} = 2\, \left ({\sigma_1}_{ij}^{-2} + {\sigma_2}_{ij}^{-2}\right )^{-1},
\label{alphageneral}
\end{eqnarray}
which is symmetric but can be non-diagonal. Expansion of the probability and the cross section into series with a small $\alpha^{-1}$
turns out to be not a Lorentz-covariant procedure, but it becomes so in the relativistic case.

\subsection{Benchmark case: $2$ wave packets $\rightarrow$ plane waves}\label{bench}

\begin{figure}
\centering
\includegraphics[width=9.00cm, height=3.80cm]{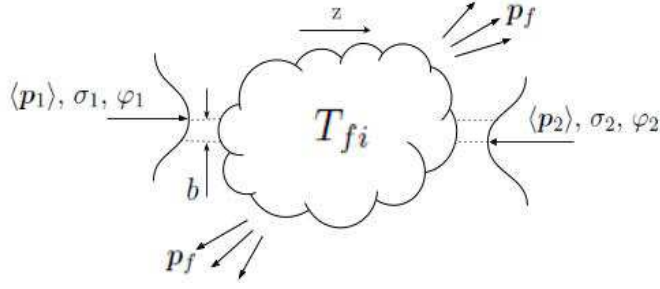}
\caption{Scattering with the in-states characterized by the mean momenta, momentum uncertainties, phases 
(will be added in Sect.4.2), and probably by the spins. Depending on the phases $\varphi_{1,2} ({\bm p}_{1,2})$, the wave front may become non-gaussian. \label{Scheme}}
\end{figure}

Let us start with collision of two wave packets with $N_f$ final plane waves (see Fig.\ref{Scheme}). Our current goal is to derive formulas for the observables
that are reduced to the customary plane-wave expressions in the corresponding limit, $\sigma_1, \sigma_2 \rightarrow 0$, and can also be generalized 
when the in-states carry phases, as well as for collisions of beams. To this end, we take first the simplest ($3+1$) D coherent states\footnote{As these states are approximate, we neglect the terms $\mathcal O(\sigma^4)$ from the very beginning; see also Sec.\ref{Alt}.} (\ref{WGaussA}).
Making use of the equality
\begin{eqnarray}
& \displaystyle
\int d^3R\, e^{-i{\bm k}{\bm R}} n ({\bm r} + {\bm R}, {\bm p}, t) = \Big (\frac{2\sqrt{\pi}}{\sigma}\Big)^3\, \exp\Big\{i {\bm k} ({\bm r} - \langle {\bm r}\rangle) - \frac{({\bm p} - \langle {\bm p}\rangle)^2}{\sigma^2} - \frac{{\bm k}^2}{(2 \sigma)^2}\Big\},
\label{RGauss}
\end{eqnarray}
we arrive at the following formula for the correlator in Eq.(\ref{M}):
\begin{eqnarray}
& \displaystyle
\mathcal L ({\bm k}, {\bm p}_{1,2}) = (2\pi)^4 \upsilon \Big(\frac{2}{\sigma_1 \sigma_2}\Big)^3\, \delta \left({\bm k} {\bm u}_2 - {\bm k} {\bm u}_1\right)
\cr & \displaystyle 
\times\exp\left\{-\left (\frac{1}{(2\sigma_1)^2} + \frac{1}{(2\sigma_2)^2}\right ) {\bm k}^2 - i{\bm k} {\bm b} -\frac{({\bm p}_1 - \langle{\bm p}\rangle_1)^2}{\sigma_1^2} -\frac{({\bm p}_2 - \langle{\bm p}\rangle_2)^2}{\sigma_2^2}\right\}
\label{L}
\end{eqnarray}
where 
$$
{\bm b} = {\bm r}_{0,1} - {\bm r}_{0,2}
$$
is the relative impact parameter of two particles at $t=0$.

The small values of ${\bm k}$ give the main contribution to the integral in (\ref{M}) and, consequently,
one can make the following expansions\footnote{We shall not deal with the non-plane-wave matrix element $S_{fi}$ any longer, 
that is why below we shall omit the superscript $(pw)$ of $T_{fi}$: $T_{fi}^{(pw)} \rightarrow T_{fi}$.},
\begin{eqnarray}
& \displaystyle
\varepsilon ({\bm p} \pm {\bm k}/2) \approx \varepsilon ({\bm p}) \pm \frac{1}{2}{\bm k}{\bm u} + \frac{1}{8\varepsilon ({\bm p})} \left(\delta_{ij} - u_i u_j \right) k_i k_j,
\cr
& \displaystyle
T_{fi} ({\bm p}_1 + {\bm k}/2, {\bm p}_2 - {\bm k}/2) T_{fi}^* ({\bm p}_1 - {\bm k}/2, {\bm p}_2 + {\bm k}/2) 
\approx |T_{fi} ({\bm p}_1, {\bm p}_2)|^2 + 
\cr & \displaystyle
+ 2i k_m \Im \{T_{fi}^* \partial_{k_m} T_{fi}\}_{{\bm k} = 0} + k_m k_n \left[-(\partial_{k_m} T_{fi})(\partial_{k_n} T_{fi}^*) + 
\Re \{T_{fi}^* \partial^2_{k_m k_n} T_{fi}\}\right]_{{\bm k} = 0} \equiv 
\cr & \displaystyle
\equiv |T_{fi} ({\bm p}_1, {\bm p}_2)|^2 + k_m C_m ({\bm p}_1, {\bm p}_2) + k_m k_n D_{mn} ({\bm p}_1, {\bm p}_2)
\label{T}
\end{eqnarray}
where $ \partial_{k_m} T_{fi} ({\bm p}_1 + {\bm k}/2, {\bm p}_2 - {\bm k}/2) = \frac{1}{2} \left (\partial_{p_{1,m}} - \partial_{p_{2,m}}\right ) T_{fi} ({\bm p}_1, {\bm p}_2)$.
We imply that the amplitude be a smooth- and analytical function of its arguments. 

In what follows we shall need the integrals
\begin{eqnarray}
& \displaystyle
\int d^3k \exp \left\{-k_i A_i -\frac{1}{2} B_{ij} k_i k_j\right\} \left\{1, k_m, k_m k_n\right\} = \frac{(2\pi)^{3/2}}{\sqrt{\det B}}\, \exp \left\{\frac{1}{2} B^{-1}_{ij} A_i A_j\right\}
\cr & \displaystyle
 \times\left\{1,\, - B^{-1}_{mi} A_i,\, B^{-1}_{m n} + B^{-1}_{mi}B^{-1}_{nj} A_iA_j\right\},
\label{Gaussian}
\end{eqnarray}
where in our case we find:
\begin{eqnarray}
& \displaystyle
B_{ij} = \alpha\, \delta_{ij} - \beta\,(\delta_{ij} - u_{1,i}u_{1,j}),\,\,\,\,\,
\beta = \frac{it}{4} \left(\frac{1}{\varepsilon_1} + \frac{1}{\varepsilon_2}\right),
\cr & \displaystyle
\det B = (\alpha - \beta)^2 (\alpha -\beta (1 - {\bm u}_1^2)),\ B^{-1}_{ij} = \frac{\alpha - \beta}{\det B} \left(\delta_{ij}\, (\alpha - \beta (1 - {\bm u}_1^2)) - u_{1,i} u_{1,j}\,\beta\right),
\cr & \displaystyle
{\bm A} = i \langle {\bm r}_1\rangle - i \langle {\bm r}_2\rangle = i \left({\bm b} + t^{\prime} \Delta {\bm u}\right),\ \Delta {\bm u} = {\bm u}_1 - {\bm u}_2,
\label{B}
\end{eqnarray}
with $\alpha$ from Eq.(\ref{alpha}). Note that by virtue of the equality ${\bm k} {\bm u}_2 = {\bm k} {\bm u}_1$, one could have written instead $u_{2,i}u_{2,j}$ in $B_{ij}$. 

Using integral representations of the delta-functions, we obtain the following expression for the integral over ${\bm k}$ in Eq.(\ref{M}):
\begin{eqnarray}
& \displaystyle
\int \frac{d^3k}{(2\pi)^3}\, \delta \Big (\varepsilon_1 ({\bm p}_1 + {\bm k}/2) + \varepsilon_2 ({\bm p}_2 - {\bm k}/2) - \varepsilon_f \Big )\, \delta \left({\bm k} {\bm u}_2 - {\bm k} {\bm u}_1\right)
\cr & \displaystyle
\times \exp\left\{-\left (\frac{1}{(2\sigma_1)^2} + \frac{1}{(2\sigma_2)^2}\right ) {\bm k}^2 - i{\bm k} {\bm b}\right\} T_{fi} ({\bm p}_1 + {\bm k}/2, {\bm p}_2 - {\bm k}/2) T_{fi}^* ({\bm p}_1 - {\bm k}/2, {\bm p}_2 + {\bm k}/2) \approx 
\cr & \displaystyle
\approx (2\pi)^{-3/2}\int \frac{dt^{\prime}}{2\pi}\frac{dt}{2\pi}\, e^{it (\varepsilon_1 ({\bm p}_1) + \varepsilon_2 ({\bm p}_2) - \varepsilon_f)} \frac{1}{\sqrt{\det B}}\, \exp\Big\{-\frac{1}{2} \left(\langle {\bm r}_1\rangle - \langle {\bm r}_2\rangle\right)_i\left(\langle {\bm r}_1\rangle - \langle {\bm r}_2\rangle\right)_j B^{-1}_{ij}\Big\} 
\cr & \displaystyle
\times \Big\{|T_{fi}|^2 - i B^{-1}_{ij} \left(\langle {\bm r}_1\rangle - \langle {\bm r}_2\rangle\right)_i C_j + [B^{-1}_{ij} - B^{-1}_{im}B^{-1}_{jn}\left(\langle {\bm r}_1\rangle - \langle {\bm r}_2\rangle\right)_m \left(\langle {\bm r}_1\rangle - \langle {\bm r}_2\rangle\right)_n]D_{ij}\Big\}
\label{k}
\end{eqnarray}
and the integral over $t^{\prime}$ is gaussian and can be easily evaluated. After some algebra, we arrive at the following probability formula:
\begin{eqnarray}
& \displaystyle
dW = \prod\limits_{f=3}^{N_f+2} \frac{d^3 p_f}{(2\pi)^3}\, \frac{(2\pi)^9}{(\pi\sigma_1 \sigma_2)^3} \int \frac{d^3 p_1}{(2\pi)^3}\frac{d^3 p_2}{(2\pi)^3} \frac{dt}{2\pi}\,\, \delta^{(3)} ({\bm p}_1 + {\bm p}_2 - {\bm p}_f)\, \frac{1}{\sqrt{\det B\, \Delta u B^{-1} \Delta u}}
\cr & \displaystyle
\times\exp\Big\{it (\varepsilon_1 ({\bm p}_1) + \varepsilon_2 ({\bm p}_2) - \varepsilon_f) -\frac{({\bm p}_1 - \langle{\bm p}\rangle_1)^2}{\sigma_1^2} -\frac{({\bm p}_2 - \langle{\bm p}\rangle_2)^2}{\sigma_2^2} - \frac{1}{2}\, b B^{-1} b + \frac{1}{2} \frac{(b B^{-1} \Delta u)^2}{\Delta u B^{-1} \Delta u}\Big\}
\cr & \displaystyle
 \times \Bigg (|T_{fi}|^2 + 2 B^{-1}_{mn} \left(b_m - \Delta u_m\, \frac{b B^{-1} \Delta u}{\Delta u B^{-1} \Delta u}\right) \Im \{T_{fi}^* \partial_{k_n} T_{fi}\}_{{\bm k} = 0} + 
\cr & \displaystyle 
+ \Big (B^{-1}_{mn} - B^{-1}_{mk}B^{-1}_{nl}\frac{\Delta u_k \Delta u_l}{\Delta u B^{-1} \Delta u} \Big )
\left[-(\partial_{k_m} T_{fi})(\partial_{k_n} T_{fi}^*) + \Re \{T_{fi}^* \partial^2_{k_m k_n} T_{fi}\}\right]_{{\bm k} = 0}\Bigg )
\label{MApp}
\end{eqnarray}
Note that the first correction to $|T_{fi}|^2$ vanishes when ${\bm b} = 0$.

An attentive reader might have already noticed that in the plane-wave regime, when $\sigma_{1} = \sigma_{2} \equiv \sigma \rightarrow 0$, the following limits hold true:
$$
\det B \rightarrow \sigma^{-6},\ \alpha \rightarrow \sigma^{-2},\ B^{-1}_{ij} \rightarrow \sigma^2 \delta_{ij},
$$
that is why we wrote down in (\ref{MApp}) only $\mathcal O(\sigma^2)$-corrections to $|T_{fi}|^2$ since we have already neglected $\mathcal O(\sigma^4)$-terms in the Wigner functions (see Eq.(\ref{WGaussAO3})). 
For vanishing $\sigma$, the probability (\ref{MApp}) decreases as
$$
\exp\{-{\bm b}^2 \sigma^2/2\}
$$

If the wave packets do not spread much during the collision (say, for ultra-relativistic neutrinos \cite{Akhmedov_10}),
that is,
\begin{eqnarray}
& \displaystyle
t_{\text{col}} \ll t_{\text{diff}} \sim \frac{1}{\sigma u_{\perp}} \sim \frac{\varepsilon}{\sigma^2}
\label{spread}
\end{eqnarray}
then one can neglect $|\beta|$ compared to $\alpha$, and the ``standard'' energy delta-function, $\delta (\varepsilon_1 ({\bm p}_1) + \varepsilon_2 ({\bm p}_2) - \varepsilon_f)$, is recovered. 
This is so, in particular, in the plane-wave regime:
\begin{eqnarray}
& \displaystyle
dW^{(pw)} = dW|_{\sigma \rightarrow 0} =(2\pi)^4\, \delta^{(4)} \left (\langle p_1\rangle + \langle p_2\rangle - p_f\right) |T_{fi}|^2\, \frac{\sigma^2}{2\pi} \frac{1}{|\Delta {\bm u}|} \prod\limits_{f=3}^{N_f+2} \frac{d^3 p_f}{(2\pi)^3}
\label{MPW}
\end{eqnarray}
where $|\Delta {\bm u}| = \sqrt{\left ({\bm u}_1 (\langle{\bm p}\rangle_1) - {\bm u}_2 (\langle{\bm p}\rangle_2)\right )^2}$ 
. If Eq.(\ref{Vrule}) is then applied, we have $\sigma^2/(2\pi) \rightarrow 1/(2 V^{2/3})$, 
and the plane-wave cross-section is obtained dividing (\ref{MPW}) by $1/(2 V^{2/3}|\Delta {\bm u}|)$ instead of $T/V$ (see Ref.\cite{BLP}). 
This implies the so-called ``time-to-space conversion''\footnote{In the plane-wave approximation, $T$ and $V$ are independent parameters, 
whereas now the time uncertainty, $\delta t \sim 1/\delta \varepsilon$, is determined by the momentum one as 
$\varepsilon \delta \varepsilon = p \sigma$. This is the reason for the ``time-to-space conversion''.} \cite{Akhmedov_09}, $T = V^{1/3}/(2 |\Delta {\bm u}|)$. 
One can avoid these (purely technical) subtleties by dealing with the cross section (\ref{Cr}) instead, which is obtained by dividing (\ref{MApp}) by the luminosity.
Within the same accuracy, the latter is
\begin{eqnarray}
& \displaystyle
L = \frac{(8\pi)^2}{\sigma_1 \sigma_2 (\sigma_1^2 + \sigma_2^2)} \int \frac{d^3 p_1}{(2\pi)^3}\frac{d^3 p_2}{(2\pi)^3} \frac{\upsilon ({\bm p}_1, {\bm p}_2)}{|\Delta {\bm u}|} 
\cr & \displaystyle
\times \exp \Big\{-\frac{\sigma_1^2 \sigma_2^2}{\sigma_1^2 + \sigma_2^2} \Big ({\bm b}^2 - \frac {(\Delta {\bm u} {\bm b})^2}{(\Delta {\bm u})^2}\Big ) - \frac{({\bm p}_1 - \langle{\bm p}\rangle_1)^2}{\sigma_1^2} -\frac{({\bm p}_2 - \langle{\bm p}\rangle_2)^2}{\sigma_2^2}\Big\}
\label{LL}
\end{eqnarray}
where the first terms in the exponent describes overlap of the incoming states. In the plane-wave limit, we have exactly the factor $L \rightarrow \sigma^2 \upsilon (\langle{\bm p}\rangle_1, \langle{\bm p}\rangle_2)/(2\pi |\Delta {\bm u}|)$ that cancels the extra term in (\ref{MPW}) and leaves us with the conventional plane-wave cross section.

We would like to emphasize that the probability formula (\ref{MApp}) 
includes plane-wave processes as a special case. Not only does this expression describe all the well-known specifically quantum phenomena, 
such as recoil- and spin-flip effects for instance, it also describes quantum effects that have no classical counterpart and \textit{vanish in the plane-wave approximation}, 
although we have used the everywhere-positive Wigner functions in the derivation. 


\subsection{$2$ packets with phases $\rightarrow$ plane waves}\label{phases}

Now let both the in-states possess phases, $\varphi_1 ({\bm p}_1)$ and $\varphi_2 ({\bm p}_2)$. 
Calculating the particle correlator in Eq.(\ref{M}) with the Wigner functions from (\ref{Wapp}), 
we note that the phase terms do not depend on the integration variables and, consequently, the final results for the correlator, 
the luminosity, and for the probability are given by the Eqs. (\ref{L}), (\ref{LL}), and (\ref{MApp}), respectively, simply with the following substitution:
\begin{eqnarray}
& \displaystyle
{\bm b} \rightarrow {\bm b}_{\varphi} = {\bm b} - \frac{\partial \varphi_1 ({\bm p}_1)}{\partial {\bm p}_1} + \frac{\partial \varphi_2 ({\bm p}_2)}{\partial {\bm p}_2}.
\label{bsubst}
\end{eqnarray}
We shall call this vector ${\bm b}_{\varphi}$ the \textit{effective} impact parameter. For the vortex- and Airy particles we have 
$$
\frac{\partial \varphi ({\bm p})}{\partial {\bm p}} = \ell\, \frac{\hat{{\bm z}} \times {\bm p}}{{\bm p}_{\perp}^2}\ \ \text{and}\ \ 
\frac{\partial \varphi ({\bm p})}{\partial {\bm p}} = {\bm \eta} = \{\xi_x^3 p_x^2, \xi_y^3 p_y^2, 0\},
$$
respectively, see Eqs.(\ref{WGaussOAM}), (\ref{WAiryapp}). An envelope determining dependence of the probability (\ref{MApp}) upon ${\bm b}_{\varphi}$ is
\begin{eqnarray}
& \displaystyle
\exp\Big\{- \frac{1}{2} b_{\varphi} B^{-1} b_{\varphi} + \frac{1}{2} \frac{( b_{\varphi} B^{-1} \Delta u)^2}{\Delta u B^{-1} \Delta u}\Big\} 
\label{terms}
\end{eqnarray}
For vortex particles with a very small $\sigma$, this is proportional to
\begin{eqnarray}
& \displaystyle
\exp\left\{- \sigma^2 \ell^2/p_{\perp}^2\right \} \equiv \exp\{-\ell^2/\ell_{\text{max}}^2\}
\label{ell}
\end{eqnarray}
with $\ell_{\text{max}}$ from Eq.(\ref{lpackmax}).


If only one of the in-states is a plane wave with $\sigma \rightarrow 0$, then
$$
B^{-1}_{ij} \rightarrow 2\sigma^2 \delta_{ij} \rightarrow 0,
$$ 
and the phase-dependent terms (\ref{terms}) vanish anyway. The probability in this case does not depend on a phase of the second (non-plane-wave) state at all.
In other words, the observables become sensitive to the in-states' phases if and only if the normalized wave-packets are used \textit{for both} of the in-states.
This observation generalizes the analogous conclusion for vortex particles \cite{I_PRD} and can be easily understood: 
dependence upon the phases appears thanks to an overlap of the in-states, which are not orthogonal.
When at least one of them is a plane-wave or even a non-normalized pure Bessel- or Airy state, then this overlap vanishes together with $\alpha^{-1}$ from Eq.(\ref{alpha}).

We return now to the correction to $|T_{fi}|^2$ in (\ref{MApp}) that is linear in ${\bm b}_{\varphi}$ 
and note that when the packets possess phases this correction \textit{survives} even when ${\bm b} = 0$. 
It depends on phases of the in-states as well as on the overall phase of the scattering amplitude, 
$\zeta_{fi} ({\bm p}_1, {\bm p}_2)$. Indeed, if one represents the amplitude as follows
\begin{eqnarray}
& \displaystyle
T_{fi} = |T_{fi}|\, \exp\left\{i \zeta_{fi}\right\},
\label{phase_2}
\end{eqnarray}
then one can obtain the following simple expression:
\begin{eqnarray}
& \displaystyle
|T_{fi}|^2 + 2 B^{-1}_{mn} \left(b_m - \Delta u_m\, \frac{b B^{-1} \Delta u}{\Delta u B^{-1} \Delta u}\right) \Im \{T_{fi}^* \partial_{k_n} T_{fi}\}_{{\bm k} = 0} =
\cr & \displaystyle
= |T_{fi}|^2 \left (1 + {\bm K}({\bm p}_1, {\bm p}_2, t) \left (\frac{\partial \zeta_{fi}}{\partial {\bm p}_1} - \frac{\partial \zeta_{fi}}{\partial {\bm p}_2}\right )\right )
\label{corr1}
\end{eqnarray}
where the function
\begin{eqnarray}
& \displaystyle
{\bm K}({\bm p}_1, {\bm p}_2, t) = \frac{1}{\alpha} \Big ({\bm b}_{\varphi} - \frac{\beta}{\alpha + \beta {\bm u}_1^2}\, {\bm u}_1 ({\bm u}_1 {\bm b}_{\varphi}) - \frac{b_{\varphi} B^{-1} \Delta u}{\Delta u B^{-1} \Delta u}\, \Delta {\bm u} + 
\cr & \displaystyle \qquad \qquad \qquad \qquad \qquad \qquad \qquad \qquad \qquad 
+ \frac{\beta}{\alpha + \beta {\bm u}_1^2}\, \frac{b_{\varphi} B^{-1} \Delta u}{\Delta u B^{-1} \Delta u}\, (\Delta {\bm u}{\bm u}_1) {\bm u}_1\Big ).
\label{K1}
\end{eqnarray}
is odd in ${\bm b}_{\varphi}$. The second correction to $|T_{fi}|^2$ in (\ref{MApp}), on the contrary, does not depend on the effective impact parameter:
\begin{eqnarray}
& \displaystyle
-(\partial_{k_m} T_{fi})(\partial_{k_n} T_{fi}^*) + \Re \{T_{fi}^*\, \partial^2_{k_m k_n} T_{fi}\} = 
\cr & \displaystyle
= |T_{fi}|\, \partial^2_{k_m k_n} |T_{fi}| - (\partial_{k_m} |T_{fi}|) (\partial_{k_n} |T_{fi}|) - 2 |T_{fi}|^2 (\partial_{k_m} \zeta_{fi}) (\partial_{k_n} \zeta_{fi})
\label{corr2}
\end{eqnarray}
where an imaginary part on the left-hand side vanishes when convoluted with a $(m, n)$-symmetric expression in (\ref{MApp}).

In order to quantify an effect of the phase $\zeta_{fi}$, we define the following asymmetry:
\begin{eqnarray}
& \displaystyle
\mathcal A[{\bm b}_{\varphi}] = \frac{dW [{\bm b}_{\varphi}] - dW[-{\bm b}_{\varphi}]}{dW [{\bm b}_{\varphi}] + dW[-{\bm b}_{\varphi}]} = \frac{d\sigma [{\bm b}_{\varphi}] - d\sigma[-{\bm b}_{\varphi}]}{d\sigma [{\bm b}_{\varphi}] + d\sigma[-{\bm b}_{\varphi}]} = \frac{d\sigma^{(1)} [{\bm b}_{\varphi}] - d\sigma^{(1)}[-{\bm b}_{\varphi}]}{2 d\sigma^{(pw)}} + \mathcal O (\sigma^4),
\label{Asymm}
\end{eqnarray}
which vanishes in the plane-wave limit. Explicit formulas for the first correction $d\sigma^{(1)}$ to the plane-wave cross section $d\sigma^{(pw)}$ and for the asymmetry 
will be given hereafter.

There are two ways how one can change the sign of ${\bm b}_{\varphi}$: 
\begin{itemize}
\item
If the incoming states are just wave packets with no phases whatsoever, 
the sign of ${\bm b}_{\varphi} = {\bm b}$ (the latter is to be small, $b \lesssim 1/\sigma$, but non-vanishing) 
can be changed by replacing the initial wave packets:
$$
{\bm r}_{0,1} \leftrightarrow {\bm r}_{0,2}
$$
\item
Conversely, when ${\bm b} = 0$ and the in-states possess phases, the change ${\bm b}_{\varphi} \rightarrow - {\bm b}_{\varphi}$ can be achieved by inverting a sign of parameters, 
such as the OAM $\ell$ or the vector ${\bm \xi}$ for Airy beams. The phases must contain only odd degrees of the parameters, which is the case for both types of the states.
\end{itemize}
It is clear that in both these scenarios we need to deal with the realistic beams of $N_b \gg 1$ particles and of a width $\sigma_b$ instead of single wave packets.

\subsection{Alternative representation of the probability formula}\label{Alt}

When deriving the general probability formula (\ref{MApp}), we used the approximate expressions for the Wigner functions and for the correlator, neglecting $\mathcal O(\sigma^4)$-corrections.
Here we show how to obtain a formula, equivalent to (\ref{MApp}) within this accuracy, but written in a more compact fashion, when the assumption of small $\sigma_{1,2}$ 
is made only once and at the very last stage\footnote{I thank A. Di Piazza for pointing this out to me.}. Such a procedure allows us to come to the final result quicker, 
however, all the properties of the Wigner functions remain hidden here. We start with the exact (i.e. not expanded over the small $\sigma$) Wigner function,
\begin{eqnarray}
& \displaystyle
n ({\bm r}, {\bm p}, t) = \frac{1}{(\sqrt{\pi}\, \sigma)^3}\int d^3k \exp\Big\{-\frac{({\bm p} - \langle {\bm p}\rangle)^2}{\sigma^2} - \frac{{\bm k}^2}{(2\sigma)^2} + i {\bm k} ({\bm r} - {\bm r}_0) 
\cr & \displaystyle
-it \left (\varepsilon ({\bm p} + {\bm k}/2) - \varepsilon ({\bm p} - {\bm k}/2)\right ) + i \left (\varphi ({\bm p} + {\bm k}/2) - \varphi ({\bm p} - {\bm k}/2)\right )\Big\},
\label{Eq1Alt}
\end{eqnarray}
and the exact correlator:
\begin{eqnarray}
& \displaystyle
\mathcal L ({\bm p}_1, {\bm p}_2, {\bm k}) = \frac{(2\pi)^7 \upsilon}{(\pi\, \sigma_1 \sigma_2)^3}\, \delta \left (\varepsilon_1 ({\bm p}_1 + {\bm k}/2) - \varepsilon_1 ({\bm p}_1 - {\bm k}/2) + \varepsilon_2 ({\bm p}_2 - {\bm k}/2) - \varepsilon_2 ({\bm p}_2 + {\bm k}/2)\right )
\cr & \displaystyle
\times \exp\Big\{-\frac{({\bm p}_1 - \langle {\bm p}\rangle_1)^2}{\sigma_1^2} - \frac{({\bm p}_2 - \langle {\bm p}\rangle_2)^2}{\sigma_2^2} - {\bm k}^2\left (\frac{1}{(2\sigma_1)^2} + \frac{1}{(2\sigma_2)^2}\right ) -
\cr & \displaystyle - i {\bm k} {\bm b} + i \left (\varphi_1 ({\bm p}_1 + {\bm k}/2) - \varphi_1 ({\bm p}_1 - {\bm k}/2) + \varphi_2 ({\bm p}_2 - {\bm k}/2) - \varphi_2 ({\bm p}_2 + {\bm k}/2)\right )\Big\}.
\label{Eq2Alt}
\end{eqnarray}
This yields the following (exact) expression for the probability:
\begin{eqnarray}
& \displaystyle
dW = \prod\limits_{f=3}^{N_f+2}\frac{d^3p_f}{(2\pi)^3}\,\frac{(2\pi)^{11}}{(\pi\, \sigma_1 \sigma_2)^3}\, \int\frac{d^3p_1}{(2\pi)^3}\frac{d^3p_2}{(2\pi)^3}\frac{d^3 k}{(2\pi)^3}\,
\delta \left (\varepsilon_1 ({\bm p}_1 - {\bm k}/2) + \varepsilon_2 ({\bm p}_2 + {\bm k}/2) - \varepsilon_f\right )
\cr & \displaystyle
\times \delta \left (\varepsilon_1 ({\bm p}_1 + {\bm k}/2) + \varepsilon_2 ({\bm p}_2 - {\bm k}/2) - \varepsilon_f\right ) \delta ({\bm p}_1 + {\bm p}_2 - {\bm p}_f)
\cr & \displaystyle 
\times T_{fi} ({\bm p}_1 + {\bm k}/2, {\bm p}_2 - {\bm k}/2) T_{fi}^*({\bm p}_1 - {\bm k}/2, {\bm p}_2 + {\bm k}/2)
\cr & \displaystyle
\times \exp\Big\{-\frac{({\bm p}_1 - \langle {\bm p}\rangle_1)^2}{\sigma_1^2} - \frac{({\bm p}_2 - \langle {\bm p}\rangle_2)^2}{\sigma_2^2} - {\bm k}^2\left (\frac{1}{(2\sigma_1)^2} + \frac{1}{(2\sigma_2)^2}\right ) -
\cr & \displaystyle - i {\bm k} {\bm b} + i \left (\varphi_1 ({\bm p}_1 + {\bm k}/2) - \varphi_1 ({\bm p}_1 - {\bm k}/2) + \varphi_2 ({\bm p}_2 - {\bm k}/2) - \varphi_2 ({\bm p}_2 + {\bm k}/2)\right )\Big\}
\label{Eq3Alt}
\end{eqnarray}

Now we expand all the functions in series over the small ${\bm k}$ up to the 2nd order inclusive, thus keeping terms not higher than $\mathcal O(\sigma^2)$,
and then integrate over ${\bm k}$, similar to the procedure in Sec.\ref{bench}.
The result is:
\begin{eqnarray}
& \displaystyle
dW = \prod\limits_{f=3}^{N_f+2}\frac{d^3p_f}{(2\pi)^3}\,\frac{(2\pi)^9}{(\pi\sigma_1 \sigma_2)^3}\, \int\frac{d^3p_1}{(2\pi)^3}\frac{d^3p_2}{(2\pi)^3}\frac{dt}{2\pi}\,
\frac{\delta ({\bm p}_1 + {\bm p}_2 - {\bm p}_f)}{\sqrt{\det B\, \Delta u B^{-1} \Delta u}}
\cr & \displaystyle 
\Bigg (|T_{fi}|^2 + \frac{1}{4} \mathcal C_{ij} \Big [B^{-1}_{ij} 
- B^{-1}_{im}B^{-1}_{jn} \Big(\tilde{b}{_{\varphi}}_m \tilde{b}{_{\varphi}}_n
- 2 \Delta u_m \tilde{b}{_{\varphi}}_n\, \frac{\Delta u B^{-1}\tilde{b}_{\varphi}}
{\Delta u B^{-1} \Delta u} + 
\cr & \displaystyle
+ \frac{\Delta u_m \Delta u_n}{\Delta u B^{-1} \Delta u} \Big (1 + \frac{(\Delta u B^{-1} \tilde{b}_{\varphi})^2}{\Delta u B^{-1} \Delta u}\Big )\Big )\Big ]\Bigg )
\exp\Big\{it (\varepsilon_1 ({\bm p}_1) + \varepsilon_2 ({\bm p}_2) - \varepsilon_f) -
\cr & \displaystyle 
- \frac{({\bm p}_1 - \langle {\bm p}\rangle_1)^2}{\sigma_1^2} - \frac{({\bm p}_2 - \langle {\bm p}\rangle_2)^2}{\sigma_2^2}
+ \frac{1}{2} \frac{(\Delta u B^{-1} \tilde{b}_{\varphi})^2}{\Delta u B^{-1} \Delta u} - \frac{1}{2}\,\tilde{b}_{\varphi} B^{-1}\tilde{b}_{\varphi} \Big\}
\label{Eq6Alt}
\end{eqnarray}
where we have used the following representation:
\begin{eqnarray}
& \displaystyle
T_{fi} ({\bm p}_1 + {\bm k}/2, {\bm p}_2 - {\bm k}/2) T_{fi}^*({\bm p}_1 - {\bm k}/2, {\bm p}_2 + {\bm k}/2) 
\approx
\cr & \displaystyle
\approx \Big (|T_{fi}|^2 + \frac{1}{4}\, k_i k_j\, \mathcal C_{ij} + \mathcal O(k^4)\Big )\,\exp\Big\{i{\bm k} \partial_{\Delta {\bm p}}\, \zeta_{fi} ({\bm p}_1, {\bm p}_2) + \mathcal O(k^3)\Big\},
\label{Eq4bAlta}
\end{eqnarray}
and we have also denoted:
\begin{eqnarray}
& \displaystyle
\partial_{\Delta {\bm p}} = \frac{\partial}{\partial {\bm p}_1} - \frac{\partial}{\partial {\bm p}_2},
\cr & \displaystyle
\mathcal C_{ij}({\bf p}_1, {\bf p}_2) = |T_{fi}| \partial_{\Delta {\bm p}_i}\partial_{\Delta {\bm p}_j}|T_{fi}| - (\partial_{\Delta {\bm p}_i}|T_{fi}|)(\partial_{\Delta {\bm p}_j}|T_{fi}|)
\cr & \displaystyle
\tilde{{\bm b}}_{\varphi}= {\bm b} - \frac{\partial \varphi_1 ({\bm p}_1)}{\partial {\bm p}_1} + \frac{\partial \varphi_2 ({\bm p}_2)}{\partial {\bm p}_2}  - \left(\frac{\partial}{\partial {\bm p}_1} - \frac{\partial}{\partial {\bm p}_2}\right) \zeta_{fi}.
\label{Eq4bAlt}
\end{eqnarray}
The matrix $B$ is still defined by the Eq.(\ref{B}). In contrast to Eq.(\ref{MApp}), in Eq.(\ref{Eq6Alt}) the phase $\zeta_{fi}$ also enters the exponent, 
and there are no terms linear in ${\bm b}$ in the pre-factor. As we have already neglected the $\mathcal O(\sigma^4)$ corrections when deriving both of these expressions, 
they can be easily shown to be equivalent within this accuracy.


\subsection{Generalization for beams}\label{beams}

The probability formula derived above describes scattering of two wave packets with the spatial widths $\sim 1/\sigma_{1,2}$.
However the real beams with $N_b$ particles are many orders of magnitude wider in the majority of cases: say, for the LHC beam $1/\sigma$ is less than a femtometer 
and $\sigma_b \sim 10 \mu$m. Taking as an example beams with the Gaussian distributions from Eq.(\ref{fG}), we perform statistical averaging of the particle correlator (\ref{Eq2Alt}).
The result is
\begin{eqnarray}
& \displaystyle
{\mathcal L}_b = N_{b,1} N_{b,2} \frac{(2\pi)^7 \upsilon}{(\pi\, \sigma_1 \sigma_2)^3}\, \delta \left (\varepsilon_1 ({\bm p}_1 + {\bm k}/2) - \varepsilon_1 ({\bm p}_1 - {\bm k}/2) + \varepsilon_2 ({\bm p}_2 - {\bm k}/2) - \varepsilon_2 ({\bm p}_2 + {\bm k}/2)\right )
\cr & \displaystyle
\times \exp\Big\{-\frac{({\bm p}_1 - \langle {\bm p}\rangle_1)^2}{\sigma_1^2} - \frac{({\bm p}_2 - \langle {\bm p}\rangle_2)^2}{\sigma_2^2} - \left(\frac{{\bm k}}{2}\right)^2\left (\frac{1}{\Sigma_1^2} + \frac{1}{\Sigma_2^2}\right ) -
\cr & \displaystyle - i {\bm k} {\bm b} + i \left (\varphi_1 ({\bm p}_1 + {\bm k}/2) - \varphi_1 ({\bm p}_1 - {\bm k}/2) + \varphi_2 ({\bm p}_2 - {\bm k}/2) - \varphi_2 ({\bm p}_2 + {\bm k}/2)\right )\Big\}.
\label{Corrbeam}
\end{eqnarray}
where the relative impact parameter of the two beams, 
$$
{\bm b} = {\bm r}_{b,1} - {\bm r}_{b,2} = \{{\bm b}_{\perp}, 0\},
$$
is introduced and 
\begin{eqnarray}
& \displaystyle
\Sigma_{1,2}^2 = \frac{\sigma_{1,2}^2}{1 + \sigma_{1,2}^2\sigma_{b,1,2}^2} \equiv \mathcal O(\sigma_b^{-2}).
\label{Sigma}
\end{eqnarray}
The corresponding number of events is
\begin{eqnarray}
dN = N_{b,1} N_{b,2}\, dW\left (\alpha (\sigma_{1,2}) \rightarrow \alpha(\Sigma_{1,2})\right ).
\label{Nbeam}
\end{eqnarray}
That is to say, one just needs to replace $\sigma_{1,2}$ in $\alpha$ from Eq.(\ref{alpha}) with $\Sigma_{1,2}$ in the probability formula (Eq.(\ref{MApp}) or (\ref{Eq6Alt})):
\begin{eqnarray}
\cr & \displaystyle \alpha^{-1} \rightarrow \left(\frac{1}{2\Sigma_1^2} + \frac{1}{2\Sigma_2^2}\right)^{-1} = \frac{2 \Sigma_1^2 \Sigma_2^2}{\Sigma_1^2 + \Sigma_2^2} = \mathcal O(\sigma_b^{-2}).
\label{alphabeam}
\end{eqnarray}

When $\sigma_b \gg 1/\sigma$, the effective values of ${\bm k}$ are $|{\bm k}| \lesssim 1/\sigma_b$, 
and it is much smaller than those for ${\bm p}$: $|{\bm p}| \lesssim \sigma$. As a result, the exponential envelope in the probability formula
looks as follows (see Eq.(\ref{terms})):
$$
\exp\left\{-{\bm b}_{\varphi}^2/(2\sigma_b^2)\right\}
$$
The maximum values of the parameters entering the phases $\varphi$ follow from the inequality:
\begin{eqnarray}
\left\langle\left|\frac{\partial \varphi}{\partial {\bm p}}\right|\right\rangle \lesssim \sigma_b
\label{phaseineq}
\end{eqnarray}
In particular, for beams with a phase vortex we come to the very same estimate (\ref{lpackmax}),
$\ell_{\text{max}} \sim p_{\perp}\sigma_{b} \sim \sigma\sigma_b$. Similar considerations for Airy beams, yield (cf. with Eq.(\ref{xiineq}))
\begin{eqnarray}
& \displaystyle
\xi_{\text{max}} \sim \sigma_b\quad \text{when}\quad \sigma\sigma_b \sim 1,\quad \text{or}\quad \xi_{\text{max}} \sim \frac{\sigma_b}{(\sigma \sigma_b)^{2/3}} \ll \sigma_b\quad \text{when}\quad \sigma\sigma_b \gg 1.
\label{Airymax}
\end{eqnarray}
Unlike the OAM's effective value for vortex beams, $\xi_{\text{max}}$ gets smaller when $\sigma\sigma_b \gg 1$.

\subsection{First correction to the plane-wave cross section}\label{Corr}

The integrals in Eqs.(\ref{MApp}), (\ref{Eq6Alt}) can be evaluated numerically for a specific model of $T_{fi}$. 
It would be much more illustrative, however, to have at hand a purely analytical model-independent expression 
for corrections to the conventional plane-wave results. 
To this end, we expand Eq.(\ref{Eq6Alt}), or rather its generalization for beams, into $\Sigma_{1,2}$-series when spreading of the packets is small (but not vanishing) 
according to Ineq.(\ref{spread}). Then we expand all the functions under the integral into ${\bm p}_{1,2} - \langle {\bm p} \rangle_{1,2}$ series, also up to the 2nd order inclusive, 
and finally integrate over ${\bm p}_{1,2}$ and $t$. In doing so, one can notice that
\begin{eqnarray}
& \displaystyle
B^{-1}_{ij} \rightarrow \frac{1}{\alpha} \Big (\delta_{ij} - u_{1,i}u_{1,j} \frac{\beta}{\alpha}\Big ),
\cr & \displaystyle
\frac{1}{\sqrt{\det B\, \Delta u B^{-1} \Delta u}} = \frac{1}{\alpha |\Delta {\bm u}|}\, \left (1 - \frac{it}{8\alpha} \left(\frac{1}{\varepsilon_1} + \frac{1}{\varepsilon_2}\right ) \left (\frac{\left [{\bm u}_1 \times {\bm u}_2\right ]^2}{(\Delta {\bm u})^2} - 2\right ) + \mathcal O(\sigma_b^{-4})\right ),
\label{Eq7}
\end{eqnarray}
and by virtue of this we obtain the following intermediate result:
\begin{eqnarray}
& \displaystyle
dN = dN_{\text{kin}} + d N_{\text{int}},
\cr & \displaystyle
dN_{\text{kin}} = N_{b,1} N_{b,2} \frac{(2\pi)^9}{(\pi\sigma_1 \sigma_2)^3}\,\frac{1}{\alpha}\,\int\frac{d^3p_1}{(2\pi)^3}\frac{d^3p_2}{(2\pi)^3}\,\,
\delta^{(4)} (p_1 + p_2 - p_f)\,\frac{|T_{fi}|^2}{|\Delta {\bm u}|}
\cr & \displaystyle 
\times \exp\Big\{-\frac{({\bm p}_1 - \langle {\bm p}\rangle_1)^2}{\sigma_1^2} - \frac{({\bm p}_2 - \langle {\bm p}\rangle_2)^2}{\sigma_2^2}\Big\} \prod\limits_{f=3}^{N_f+2}\frac{d^3p_f}{(2\pi)^3},
\cr & \displaystyle 
d N_{\text{int}} = N_{b,1} N_{b,2}\frac{(2\pi)^9}{(\pi\sigma_1 \sigma_2)^3}\, \frac{1}{2\alpha^2}\, \int\frac{d^3p_1}{(2\pi)^3}\frac{d^3p_2}{(2\pi)^3}\frac{dt}{2\pi}\,\,
\delta ({\bm p}_1 + {\bm p}_2 - {\bm p}_f)\frac{1}{|\Delta {\bm u}|}
\cr & \displaystyle 
\times \exp\Big\{it (\varepsilon_1 ({\bm p}_1) + \varepsilon_2 ({\bm p}_2) - \varepsilon_f) -\frac{({\bm p}_1 - \langle {\bm p}\rangle_1)^2}{\sigma_1^2} - \frac{({\bm p}_2 - \langle {\bm p}\rangle_2)^2}{\sigma_2^2}\Big\} 
\cr & \displaystyle 
\times \Bigg (|T_{fi}|^2 \Big [\frac{it}{4} \Big(\frac{1}{\varepsilon_1} + \frac{1}{\varepsilon_2}\Big) \Big (2 -\frac{\left [{\bm u}_1 \times {\bm u}_2\right ]^2}{(\Delta {\bm u})^2}\Big ) - \Big [\frac{\Delta {\bm u}}{|\Delta {\bm u}|} \times \tilde{{\bm b}}_{\varphi}\Big ]^2 \Big ] +
\cr & \displaystyle 
\qquad \qquad \qquad \qquad + \frac{1}{2}\mathcal C_{ij} \Big (\delta_{ij} - \frac{\Delta u_i \Delta u_j}{(\Delta {\bm u})^2}\Big ) \Bigg ) \prod\limits_{f=3}^{N_f+2}\frac{d^3p_f}{(2\pi)^3}
\label{Eq8}
\end{eqnarray}
where $dN_{\text{kin}}({\sigma_{1,2} \rightarrow 0}) = dN^{(pw)}$ and the second term, $d N_{\text{int}}$, 
describes \textit{interference} of the incoming packets.

Neglecting the higher-order terms, we can simply take the integrand in $d N_{\text{int}}$ in the points $\langle {\bm p}\rangle_{1,2}$.
In doing so, however, we need to be cautious since:
\begin{itemize}
\item
The functions $\partial \varphi_{1,2}/\partial {\bm p}_{1,2}|_{{\bm p}_{1,2} = \langle {\bm p}\rangle_{1,2}}$ may not be analytical everywhere or, in other words,
in the expansion $\exp\left\{-{\bm b}_{\varphi}^2/(2\sigma_b^2)\right\} \approx 1 - {\bm b}_{\varphi}^2/(2\sigma_b^2)$
the ratio ${\bm b}_{\varphi}^2/(2\sigma_b^2)$ may not be small in the entire ${\bm p}$ domain.
Say, for vortex beams it is big when $p_{\perp} \ll \ell/\sigma_b$.
The probability itself, however, is exponentially suppressed in this case and that is why one can still use the expansion but keep in mind that $p_{\perp} > \ell/\sigma_b$.
The neglected small momenta contribute to the higher powers of $\ell/(\sigma \sigma_b) < 1$ (or to $dN^{(n)}, n \geq 2$) only.
\item
Parameters of the functions $\varphi_{1,2}$ can also depend on $\sigma$ and $\sigma_b$, which is the case, say, for Airy beams with $\xi_{\text{max}} = \xi_{\text{max}} (\sigma, \sigma_b)$. 
As a result, for vanishing transverse momentum the higher-order $\sigma$-corrections to $\varphi_{1,2}$ may give contributions to the leading order:
$\xi_{x,y}^3 \langle p_{x,y}^2 \rangle = \xi_{x,y}^3 \langle p_{x,y} \rangle^2 + \xi_{x,y}^3\sigma^2/2$. When $\langle p_{x,y} \rangle \rightarrow 0$,
the seconds term survives.
\end{itemize}
Thus when making the small-$\Delta {\bm p}$ expansion, one should either suppose that the azimuthal asymmetry is broken from the very beginning, i.e. $\langle{\bm u}\rangle_{\perp, 1,2} \ne 0$, 
or keep the higher $\sigma$-terms in $\varphi_{1,2}$ so that we can always return to the special case of a vanishing transverse momentum. 
It is the latter scenario in which the mean value $\langle{\bm b}_{\varphi}\rangle$, i.e. ${\bm b}_{\varphi}$ weighted with the Gaussians
$(\sqrt{\pi}\sigma_{1,2})^{-3}\,\exp\{-({\bm p}_{1,2} - \langle {\bm p}\rangle_{1,2})^2/\sigma_{1,2}^2\}$, comes into play.

After the integration, we arrive at the following result for the first correction:
\begin{eqnarray}
& \displaystyle
dN = dN^{(pw)}+dN^{(1)},
\cr & \displaystyle
dN^{(pw)} = N_{b,1} N_{b,2}\frac{1}{\alpha}\,(2\pi)^3\, \delta^{(4)}\left (\langle p\rangle_1 + \langle p\rangle_2 - p_f \right )
\frac{|T_{fi}|^2}{|\Delta {\bm u}|}\,\prod\limits_{f=3}^{N_f+2}\frac{d^3p_f}{(2\pi)^3},
\cr & \displaystyle 
\frac{dN^{(1)}}{dN^{(pw)}} = - \frac{3}{4}\left (\frac{\sigma_1^2}{\varepsilon_1^2} \left (1 - {\bm u}_1^2\right ) + \frac{\sigma_2^2}{\varepsilon_2^2} \left (1 - {\bm u}_2^2\right )\right )  - 
\cr & \displaystyle 
- \frac{1}{2 \alpha} \Bigg (\frac{1}{8}\,\frac{|\Delta {\bm u}|}{|T_{fi}|^2}\,\left (\partial_{\varepsilon_1} + \partial_{\varepsilon_2}\right ) \Big (\frac{1}{\varepsilon_1} + \frac{1}{\varepsilon_2}\Big )
\left (2 - \frac{\left [{\bm u}_1 \times {\bm u}_2\right ]^2}{(\Delta {\bm u})^2}\right ) \frac{|T_{fi}|^2}{|\Delta {\bm u}|}  
- 
\cr & \displaystyle 
- \frac{1}{2 |T_{fi}|^2}\, \mathcal C_{ij} \Big (\delta_{ij} - \frac{\Delta u_i \Delta u_j}{(\Delta {\bm u})^2}\Big ) 
+ \Big \langle\Big [\tilde{{\bm b}}_{\varphi} \times \frac{\Delta {\bm u}}{|\Delta {\bm u}|}\Big ]^2 \Big \rangle\Bigg ). 
\label{Eq11}
\end{eqnarray}
with $\mathcal C_{ij}$ and $\tilde{{\bm b}}_{\varphi}$ from Eq.(\ref{Eq4bAlt}) and
${\bm p}_{1,2}=\langle {\bm p}\rangle_{1,2}$ is implied everywhere. 

As a last step, we can also write down the first corrections to the luminosity and to the cross section, which are supposed to be small:
\begin{eqnarray}
& \displaystyle
L = L^{(pw)} + L^{(1)},\ d\sigma = \frac{dN}{L} = d\sigma^{(pw)} + d\sigma^{(1)},\ L^{(1)} \ll L^{(pw)},\, d\sigma^{(1)} \ll d\sigma^{(pw)},
\cr & \displaystyle
L^{(pw)} = \frac{1}{\alpha} \frac{\upsilon}{2\pi} \frac{1}{|\Delta {\bm u}|},
\cr & \displaystyle
d\sigma^{(pw)} = \frac{dN^{(pw)}}{L^{(pw)}} = N_{b,1} N_{b,2}\,(2\pi)^4 \delta^{(4)} (\langle p\rangle_1 + \langle p\rangle_2 - p_f) \frac{|T_{fi}|^2}{\upsilon}\prod\limits_{f=3}^{N_f+2}\frac{d^3p_f}{(2\pi)^3},
\label{Eqa13}
\end{eqnarray}
The former is
\begin{eqnarray}
& \displaystyle
L^{(1)} = \frac{1}{2 \alpha} \frac{1}{2\pi} \Bigg (\sigma_1^2 \frac{1}{2}\, \frac{\partial^2}{\partial {\bm p}_1^2} \frac{\upsilon}{|\Delta {\bm u}|} + \sigma_2^2 \frac{1}{2}\, \frac{\partial^2}{\partial {\bm p}_2^2} \frac{\upsilon}{|\Delta {\bm u}|} 
- \frac{1}{\alpha_0} \frac{\upsilon}{|\Delta {\bm u}|} \left\langle \left[{\bm b}_{\varphi} \times \frac{\Delta {\bm u}}{|\Delta {\bm u}|}\right]^2\right\rangle\Bigg ),
\label{Eq13}
\end{eqnarray}
and everywhere ${\bm p}_{1,2}=\langle {\bm p}\rangle_{1,2}$ is implied. This expression is even in ${\bm b}_{\varphi}$, it also contains ``kinetic'' terms 
that are due to finite sizes of the packets and an interference term, proportional to $\alpha^{-1}$ and depending on the particles' phases.
The analogous correction to the cross section is
\begin{eqnarray}
& \displaystyle
d\sigma^{(1)} = d\sigma^{(pw)} \left(\frac{dN^{(1)}}{dN^{(pw)}} - \frac{L^{(1)}}{L^{(pw)}}\right),
\cr & \displaystyle 
\frac{d\sigma^{(1)}}{d\sigma^{(pw)}} = - \frac{\sigma_1^2}{\varepsilon_1^2} \frac{1}{4} \left (3 (1-{\bm u}_1^2) + \varepsilon_1^2 \frac{|\Delta {\bm u}|}{\upsilon} \frac{\partial^2}{\partial {\bm p}_1^2} \frac{\upsilon}{|\Delta {\bm u}|}\right ) - \frac{\sigma_2^2}{\varepsilon_2^2} \frac{1}{4} \left (3 (1-{\bm u}_2^2) + \varepsilon_2^2 \frac{|\Delta {\bm u}|}{\upsilon} \frac{\partial^2}{\partial {\bm p}_2^2} \frac{\upsilon}{|\Delta {\bm u}|}\right ) -
\cr & \displaystyle 
- \frac{1}{2\alpha} \Bigg (\frac{1}{8} \frac{|\Delta {\bm u}|}{|T_{fi}|^2} \left (\partial_{\varepsilon_1} + \partial_{\varepsilon_2}\right ) \Big (\frac{1}{\varepsilon_1} + \frac{1}{\varepsilon_2}\Big )
\left (2 - \frac{\left [{\bm u}_1 \times {\bm u}_2\right ]^2}{(\Delta {\bm u})^2}\right ) \frac{|T_{fi}|^2}{|\Delta {\bm u}|} -
\cr & \displaystyle 
-\frac{1}{2 |T_{fi}|^2}\, \mathcal C_{ij} \Big (\delta_{ij} - \frac{\Delta u_i \Delta u_j}{(\Delta {\bm u})^2}\Big ) + 2 \left [\frac{\Delta {\bm u}}{|\Delta {\bm u}|} \times \left [\frac{\Delta {\bm u}}{|\Delta {\bm u}|} \times \langle {\bm b}_{\varphi}\rangle \right ]\right ]\cdot \partial_{\Delta {\bm p}} \zeta_{fi} 
+ \left [\frac{\Delta {\bm u}}{|\Delta {\bm u}|} \times \partial_{\Delta {\bm p}} \zeta_{fi}\right ]^2 \Bigg ),
\label{Eqb13}
\end{eqnarray}
Here, as distinct from the number of events, dependence on the effective impact-parameter survives in the linear terms only, 
and when $\zeta_{fi} = 0$ (in some models on a tree-level, for instance) this correction depends neither on the impact-parameter nor on the particles' phases.
Note that the term linear in ${\bm b}_{\varphi}, \partial_{\Delta {\bm p}} \zeta_{fi}$
can be either positive or negative and this brings about the non-vanishing scattering asymmetry defined in Eq.(\ref{Asymm}).
As we shall see, this term also breaks an up-down symmetry in angular distributions of the final particles.



\subsection{Relativistic generalizations}\label{LorSc}

Now let us return to an arbitrary frame of reference with $\sigma \rightarrow \sigma_{ij}$ (recal Eq.(\ref{sigmatr})).
Formulas for the correlator, Eqs.(\ref{L}) and (\ref{Eq2Alt}), and for the probability, Eqs.(\ref{MApp}) and (\ref{Eq6Alt}),
stay valid with the following substitutions:
\begin{eqnarray}
& \displaystyle 
\frac{1}{\sigma_1^3 \sigma_2^3} \exp\Big\{-\frac{({\bm p}_1 - \langle{\bm p}_1\rangle)^2}{\sigma_1^2} -\frac{({\bm p}_2 - \langle{\bm p}_2\rangle)^2}{\sigma_2^2} - 
\frac{{\bm k}^2}{2} \left (\frac{1}{2\sigma_1^2} + \frac{1}{2\sigma_2^2}\right )\Big\} \rightarrow \cr
& \displaystyle 
\rightarrow \frac{1}{\det \sigma_1 \det\sigma_2} \exp\Big\{-({\bm p}_1 - \langle{\bm p}_1\rangle)\sigma_1^{-2}({\bm p}_1 - \langle{\bm p}_1\rangle) -({\bm p}_2 - \langle{\bm p}_2\rangle)\sigma_2^{-2}({\bm p}_2 - \langle{\bm p}_2\rangle) - 
\cr
& \displaystyle
- \frac{1}{2}{\bm k} \alpha {\bm k}\Big\},
\label{Lor1}
\end{eqnarray}
and with the following matrix $\alpha$ instead of the scalar from Eq.(\ref{alpha}):
\begin{eqnarray}
& \displaystyle 
{\alpha}_{ij} = \frac{1}{2}\, \left ({\sigma_1}_{ij}^{-2} + {\sigma_2}_{ij}^{-2}\right ),\ \text{or for a beam:}\  {\alpha}_{ij} = \frac{1}{2}\, \left ({\Sigma_1}_{ij}^{-2} + {\Sigma_2}_{ij}^{-2}\right ),
\label{Lor2}
\end{eqnarray}
where $\Sigma_{ij}$ is from Eq.(\ref{Sigmaij}). Then for a new matrix $B$ we find:
\begin{eqnarray}
& \displaystyle 
B^{-1}_{ij} = {\alpha}_{ij}^{-1} + \beta {\alpha}_{ij}^{-2} - \beta {\alpha}_{ik}^{-1}u_k {\alpha}_{jm}^{-1}u_m + \mathcal O (\alpha^{-3}), \cr
& \displaystyle 
\det B = \det \alpha \left (1 - \beta \Tr \alpha^{-1} + \beta {\bm u}_1 \alpha^{-1} {\bm u}_1 \right ) + \mathcal O (\alpha)
\label{Lor3}
\end{eqnarray}
with $\beta$ from Eq.(\ref{B}). As a result,
\begin{eqnarray}
& \displaystyle 
\frac{1}{\sqrt{\det B\, \Delta u B^{-1} \Delta u}} = \frac{1}{\sqrt{\det \alpha\, \Delta u \alpha^{-1} \Delta u}}\, \Bigg (1 + \frac{\beta}{2} \Big [\Tr \alpha_{\perp}^{-1} - \frac{1}{2} {\bm u}_1 \alpha_{\perp}^{-1} {\bm u}_1 - \frac{1}{2} {\bm u}_2 \alpha_{\perp}^{-1} {\bm u}_2 \Big ] + \cr
& \displaystyle 
\qquad \qquad\qquad\qquad\qquad\qquad \qquad\qquad\qquad\qquad\qquad\qquad\qquad\qquad\qquad
+ \mathcal O (\alpha^{-2})\Bigg )
\label{Lor4}
\end{eqnarray}
where
\begin{eqnarray}
& \displaystyle
{\alpha_{\perp}}_{ij}^{-1} = {\alpha}_{ij}^{-1} -{\alpha}_{ik}^{-1}{\alpha}_{jm}^{-1}  \frac{\Delta u_k \Delta u_m}{\Delta u \alpha^{-1} \Delta u}. 
\label{alphaperp}
\end{eqnarray}
is denoted. 

As neither the small ``parameter'' of this expansion, $\alpha^{-1}$, nor the condition of small spreading (\ref{spread}) is Lorentz-invariant (actually the latter inequality is automatically fulfilled for ultrarelativistic particles), the resultant first correction to the number of events, unlike its plane-wave counterpart, turns out to be non-invariant either:
\begin{eqnarray}
& \displaystyle
dN^{(pw)} = N_{b,1} N_{b,2} \frac{1}{\sqrt{\det \alpha\, \Delta u \alpha^{-1} \Delta u}}\,(2\pi)^3\, \delta^{(4)}\left (\langle p\rangle_1 + \langle p\rangle_2 - p_f \right )
|T_{fi}|^2\,\prod\limits_{f=3}^{N_f+2}\frac{d^3p_f}{(2\pi)^3},
\cr & \displaystyle 
\frac{dN^{(1)}}{dN^{(pw)}} = - \frac{1}{4}\left (\frac{\Tr \sigma_1^2 - 3 {\bm u}_1\sigma_1^2 {\bm u}_1}{\varepsilon_1^2} + \frac{\Tr \sigma_2^2 - 3 {\bm u}_2\sigma_2^2 {\bm u}_2}{\varepsilon_2^2}\right )  - 
\cr & \displaystyle 
- \frac{1}{2} \Bigg (\frac{1}{8}\,\frac{\sqrt{\Delta u \alpha^{-1} \Delta u}}{|T_{fi}|^2}\,\left (\partial_{\varepsilon_1} + \partial_{\varepsilon_2}\right ) \Big (\frac{1}{\varepsilon_1} + \frac{1}{\varepsilon_2}\Big ) \Big (\Tr \alpha_{\perp}^{-1} - \frac{1}{2} {\bm u}_1 \alpha_{\perp}^{-1} {\bm u}_1 - \frac{1}{2} {\bm u}_2 \alpha_{\perp}^{-1} {\bm u}_2 \Big )
\cr & \displaystyle \times \frac{|T_{fi}|^2}{\sqrt{\Delta u \alpha^{-1} \Delta u}}  - \frac{1}{2 |T_{fi}|^2}\, \mathcal C_{ij}\, {\alpha_{\perp}}_{ij}^{-1} + \left \langle\tilde{{\bm b}}_{\varphi}{\alpha_{\perp}}^{-1}\tilde{{\bm b}}_{\varphi} \right \rangle\Bigg ), 
\label{NLor}
\end{eqnarray}
where ${\bm p}_{1,2} = \langle {\bm p}\rangle_{1,2}$ is to be put. The terms neglected in Eq.(\ref{Lor4}) are $\mathcal O(\gamma^{-4})$ and that is why Lorentz invariance is restored in the relativistic case. 
Indeed, let in a frame where the packets are at rest on average $\sigma_x \sim \sigma_y \sim \sigma_z \equiv \sigma$.
Then in a collider frame of reference with 
\begin{eqnarray}
& \displaystyle
\langle{\bm p}\rangle_1 = - \langle{\bm p}\rangle_2 \equiv {\bm p} = {\bm u} \varepsilon = \{0,0,p\},\ \Delta {\bm u} = 2{\bm u},\ \upsilon = |\Delta {\bm u}| 
\label{frame}
\end{eqnarray}
we have:
\begin{eqnarray}
& \displaystyle
\frac{dN^{(1)}}{dN^{(pw)}} = \frac{1}{2}\frac{\sigma_1^2}{m_1^2} + \frac{1}{2}\frac{\sigma_2^2}{m_2^2} - \frac{1}{2} \left \langle{\bm b}_{\varphi}\alpha_{\perp}^{-1}{\bm b}_{\varphi} \right \rangle + 
\cr & \displaystyle
+ \frac{1}{4 |T_{fi}|^2}\,\mathcal C_{ij}\, {\alpha_{\perp}}_{ij}^{-1} + \langle {\bm b}_{\varphi}\rangle\alpha_{\perp}^{-1}\partial_{\Delta {\bm p}}\zeta_{fi} 
- \frac{1}{2}\, \partial_{\Delta {\bm p}}\zeta_{fi} \alpha_{\perp}^{-1}\partial_{\Delta {\bm p}}\zeta_{fi} + \mathcal O(\gamma^{-2}). 
\label{NLorrel}
\end{eqnarray}
which is invariant under boosts along the collision axis and where spreading of the packets is neglected. 
The (diagonal) matrix $\alpha_{\perp}^{-1}$ in this frame is
\begin{eqnarray}
& \displaystyle
\alpha_{\perp}^{-1} = \diag \left\{\frac{2\Sigma_{1,x}^2\Sigma_{2,x}^2}{\Sigma_{1,x}^2 + \Sigma_{2,x}^2}, \frac{2\Sigma_{1,y}^2\Sigma_{2,y}^2}{\Sigma_{1,y}^2 + \Sigma_{2,y}^2}, 0\right\} = \text{inv}.
\label{alphaperp}
\end{eqnarray}
Note that the corrections due to the finite sizes of the wave packets, $\sigma^2/m^2$, are positive.

The corresponding expressions for the luminosity are
\begin{eqnarray}
& \displaystyle
L^{(pw)} = \frac{1}{\sqrt{\det\alpha\, \Delta u \alpha^{-1} \Delta u}}\, \frac{\upsilon}{2\pi},\cr
& \displaystyle L^{(1)} = \frac{1}{2 \sqrt{\det\alpha}} \frac{1}{2\pi} \Bigg ( \frac{1}{2}\, \partial_{{\bm p}_1}\sigma_1^2\partial_{{\bm p}_1} \frac{\upsilon}{\sqrt{\Delta u \alpha^{-1} \Delta u}} + \frac{1}{2}\, \partial_{{\bm p}_2}\sigma_2^2\partial_{{\bm p}_2} \frac{\upsilon}{\sqrt{\Delta u \alpha^{-1} \Delta u}} 
- \cr 
& \displaystyle - \frac{\upsilon}{\sqrt{\Delta u \alpha^{-1} \Delta u}}\, \langle{\bm b}_{\varphi}{\alpha_{\perp}}^{-1} {\bm b}_{\varphi}\rangle\Bigg ),
\label{LumLor}
\end{eqnarray}
and for the cross section we obtain the following formula:
\begin{eqnarray}
& \displaystyle
\frac{d\sigma^{(1)}}{d\sigma^{(pw)}} = - \frac{1}{4}\left (\frac{\Tr \sigma_1^2 - 3 {\bm u}_1\sigma_1^2 {\bm u}_1}{\varepsilon_1^2} + \frac{\sqrt{\Delta u \alpha^{-1} \Delta u}}{\upsilon}\, \partial_{{\bm p}_1}\sigma_1^2\partial_{{\bm p}_1} \frac{\upsilon}{\sqrt{\Delta u \alpha^{-1} \Delta u}} + (1\rightarrow 2)\right )  - 
\cr & \displaystyle 
- \frac{1}{2} \Bigg (\frac{1}{8}\,\frac{\sqrt{\Delta u \alpha^{-1} \Delta u}}{|T_{fi}|^2}\,\left (\partial_{\varepsilon_1} + \partial_{\varepsilon_2}\right ) \Big (\frac{1}{\varepsilon_1} + \frac{1}{\varepsilon_2}\Big ) \Big (\Tr \alpha_{\perp}^{-1} - \frac{1}{2} {\bm u}_1 \alpha_{\perp}^{-1} {\bm u}_1 - \frac{1}{2} {\bm u}_2 \alpha_{\perp}^{-1} {\bm u}_2 \Big )\cr 
& \displaystyle \times \frac{|T_{fi}|^2}{\sqrt{\Delta u \alpha^{-1} \Delta u}} - \frac{1}{2 |T_{fi}|^2}\, \mathcal C_{ij}\, {\alpha_{\perp}}_{ij}^{-1} -2 \langle {\bm b}_{\varphi}\rangle\alpha_{\perp}^{-1}\partial_{\Delta {\bm p}}\zeta_{fi} 
+ \partial_{\Delta {\bm p}}\zeta_{fi} \alpha_{\perp}^{-1}\partial_{\Delta {\bm p}}\zeta_{fi}\Bigg ),
\label{crossLor}
\end{eqnarray}
or for relativistic energies in the collider frame (\ref{frame}):
\begin{eqnarray}
& \displaystyle
\frac{d\sigma^{(1)}}{d\sigma^{(pw)}} = \frac{1}{2}\frac{\sigma_1^2}{m_1^2} + \frac{1}{2}\frac{\sigma_2^2}{m_2^2} - \frac{1}{4}\, \partial_{{\bm p}_1}\sigma_1^2\partial_{{\bm p}_1} \frac{\upsilon}{\Delta u_z} - \frac{1}{4}\, \partial_{{\bm p}_2}\sigma_2^2\partial_{{\bm p}_2} \frac{\upsilon}{\Delta u_z} + 
\cr & \displaystyle 
+ \frac{1}{4 |T_{fi}|^2}\, \mathcal C_{ij}\, {\alpha_{\perp}}_{ij}^{-1} + \langle {\bm b}_{\varphi}\rangle\alpha_{\perp}^{-1}\partial_{\Delta {\bm p}}\zeta_{fi} 
- \frac{1}{2}\, \partial_{\Delta {\bm p}}\zeta_{fi} \alpha_{\perp}^{-1}\partial_{\Delta {\bm p}}\zeta_{fi} + \mathcal O(\gamma^{-2}).
\label{crossLorrel}
\end{eqnarray}
which is also $z$-invariant and $d\sigma^{(pw)}$ is from Eq.(\ref{Eqa13}). The first rows in (\ref{NLorrel}) and (\ref{crossLorrel}) contain ``kinematic'' terms due to the finite width of the packets,
while the second ones depend on derivatives of the amplitude in $\mathcal C_{ij}$ and on the phase $\zeta_{fi}$.
Note that the term $\langle{\bm b}_{\varphi}{\alpha_{\perp}}^{-1}{\bm b}_{\varphi} \rangle$ is absent in $d\sigma^{(1)}/d\sigma^{(pw)}$ 
and that is why when $\zeta_{fi} = 0$ the first correction to the cross section, unlike the one to the number of events (\ref{NLorrel}), 
depends neither on the impact-parameter nor on the phases of the incoming particles. 
In other words, the effective cross section turns out to be less sensitive (than the number of events) to the possible spatial inhomogeneity of the colliding wave fronts.

\subsection{QED example: $ee \rightarrow e^{\prime}e^{\prime}$}\label{Ex}

In order to illustrate the non-plane-wave effects, let us take a head-on elastic scattering of electrons in quantum electrodynamics. 
The tree-level amplitude \cite{BLP}, 
\begin{eqnarray}
& \displaystyle
M_{fi} = 4\pi e^2 \left (\frac{1}{t}\, (\bar{u}_2^{\prime}\gamma^{\mu}u_2)(\bar{u}_1^{\prime}\gamma_{\mu}u_1)-\frac{1}{u}\, (\bar{u}_1^{\prime}\gamma^{\nu}u_2)(\bar{u}_2^{\prime}\gamma_{\nu}u_1)\right )
\label{MMoeller}
\end{eqnarray}
has only an arbitrary constant phase due to the electrons' spinors $u ({\bm p})$, and that is why one can put $\zeta_{fi} = 0$. 
We shall work in the collider frame (\ref{frame}) with the identical incoming beams of the width $\sigma_b \approx 1/\sigma$. 
The ``kinematic'' term in (\ref{NLorrel}) is
\begin{eqnarray}
& \displaystyle
\frac{\sigma^2}{m^2} \approx \left (\frac{\lambda_c}{\sigma_b}\right )^2,
\label{kincorr}
\end{eqnarray}
where $\lambda_c = 1/m \equiv \hbar/(mc)$ is the electron's Compton wave length.
Even for the electrons focused to a spot of an Angstrom size \cite{Angstrom} this ratio is of the order of $10^{-6}$.

The main correction to the plane-wave result therefore is
\begin{eqnarray}
& \displaystyle
\frac{1}{4 |T_{fi}|^2}\, \mathcal C_{ij}\, {\alpha_{\perp}}_{ij}^{-1} = \frac{2}{\sigma_b^2} \frac{1}{|M_{fi}|^2} \frac{t u}{s - 4 m^2} \left (\frac{\partial^2 |M_{fi}|^2}{\partial t^2} - \frac{1}{|M_{fi}|^2} \left (\frac{\partial |M_{fi}|^2}{\partial t}\right )^2\right )
\label{Cij}
\end{eqnarray}
with $s = 4 \varepsilon^2, -t = 4 {\bm p}^2 \sin^2 (\theta_{sc}/2), u = 4 m^2 - s - t$. For the small $t$, we find
\begin{eqnarray}
& \displaystyle
\frac{1}{4 |T_{fi}|^2}\, \mathcal C_{ij}\, {\alpha_{\perp}}_{ij}^{-1}\approx -\frac{4}{\sigma_b^2\, t} + \mathcal O (t^0), 
\label{Cijsmall}
\end{eqnarray}
where the leading term,
\begin{eqnarray}
& \displaystyle
-\frac{4}{\sigma_b^2\, t} 
= \left (\frac{\lambda_c}{\sigma_b}\right )^2 \frac{4 m^2}{-t}, 
\label{Cijrel}
\end{eqnarray}
must be small by definition, although on the tree-level $-t < 4 m^2$. 
Thanks to the latter inequality, this correction can be much larger than (\ref{kincorr}) for the following scattering angles:
\begin{eqnarray}
& \displaystyle
\theta_{sc} \gtrsim \frac{\lambda_c}{\sigma_b}\,\frac{2m}{p} = \frac{2}{p\sigma_b}.
\label{smalltheta}
\end{eqnarray}
For the same electrons with the kinetic energy of $300$ keV from the Ref.\cite{Angstrom} (regardless of the OAM), we get
\begin{eqnarray}
& \displaystyle
\theta_{sc} \gtrsim 0.1^{\circ}.
\label{estimtheta}
\end{eqnarray}
Measuring angular distributions of the scattered electrons at such angles is definitely challenging although not impossible. 

\begin{figure}
\center
\includegraphics[width=15.70cm, height=5.00cm]{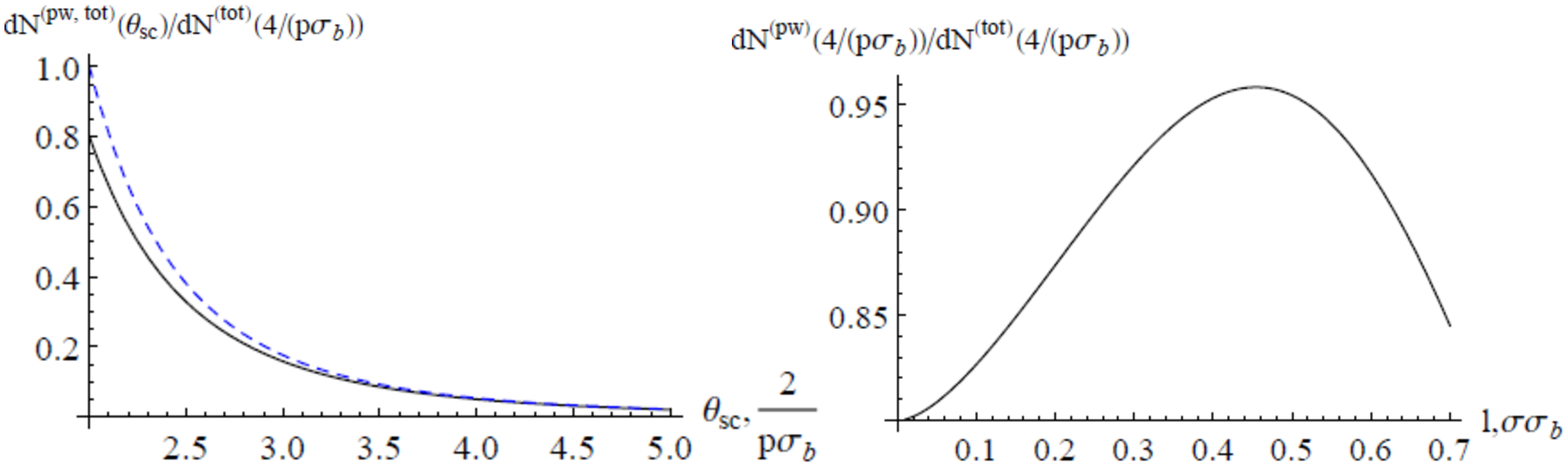}
\caption{Angular distributions of the final electrons in a tree-level M\o ller scattering with $\zeta_{fi} = 0, \varepsilon_{\text{kin}} = 300$ keV. 
\textit{Left panel}: the ordinary gaussian beams with $\sigma_b \approx 1/\sigma, {\bm b}_{\varphi} = 0$; the black solid line: the conventional plane-wave result, 
the blue dashed line: the one with the non-plane-wave corrections taken into account. 
The results are normalized to $dN^{(\text{tot})} = dN^{(pw)} + dN^{(1)}$ at $\theta_{sc} = 4/(p\sigma_b) \approx 0.2^{\circ}$.
\textit{Right panel}: Scattering of the vortex electrons with $\ell_1 = \ell_2 \equiv \ell,\, \langle {\bm p}\rangle_{\perp} = {\bm b} = 0$ and $\sigma_1 = \sigma_2 = \sigma$. 
\label{Fig_Moell}}
\end{figure}

When the incoming states possess phases the number of events also depends on the effective impact-parameter:
\begin{eqnarray}
& \displaystyle
\frac{dN^{(1)}}{dN^{(pw)}} \propto -\frac{1}{2}\langle{\bm b}_{\varphi}\alpha_{\perp}^{-1}{\bm b}_{\varphi} \rangle \approx -\frac{\langle {\bm b}_{\varphi}^2\rangle}{2\sigma_b^2}
\label{bcorr}
\end{eqnarray}
When we collide two Bessel beams with $\ell_1 = \ell_2 \equiv \ell,\, \langle {\bm p}\rangle_{\perp} = 0$ and $\sigma_1 = \sigma_2 = \sigma$, this correction becomes
\begin{eqnarray}
& \displaystyle
-\frac{\langle {\bm b}_{\varphi}^2\rangle}{2\sigma_b^2}\approx \left (\frac{\ell}{\sigma\sigma_b}\right )^2 \left (\gamma_0 + 2 \ln \left (\frac{\ell}{\sigma\sigma_b}\right ) \right ) + \mathcal O \left (\left (\frac{\ell}{\sigma\sigma_b}\right )^4 \right ),
\label{bBess}
\end{eqnarray}
which is mostly negative as $\ell < \ell_{\text{max}} \approx \sigma \sigma_b$ and where $\gamma_0 \approx 0.577$ is the Euler's constant. Calculating the mean value of ${\bm b}_{\varphi}^2$ from Eq.(\ref{bsubst}) with ${\bm b} = 0$ 
one should recall that $p_{\perp} > \ell/\sigma_b$ (see the discussion before Eq.(\ref{Eq11})).

Whereas the correction (\ref{Cij}) increases the number of events, the second one, Eq.(\ref{bBess}), diminishes it. 
As a result, these two contributions can nearly compensate each other. 
In Fig.\ref{Fig_Moell} we show angular distributions of the scattered electrons with these corrections taken into account.
As can be seen on the left panel, in a region where the correction is big but the perturbation theory still works,
its contribution can reach the values of $10-20 \%$. 

On the tree-level, neither of the corrections depend on the azimuthal angle; 
however the amplitude's finite phase, $\zeta_{fi}$, as we shall demonstrate hereafter, restores this dependence.

\section{Effects of the amplitude's phase}\label{asymm}

\subsection{Scattering asymmetry}

When the phase $\zeta_{fi}$ is non-vanishing, which is true on the loop level in QED or in the more sophisticated theories like quantum chromodynamics,
the cross section also depends on signs of the incoming particles' phases (say, on OAM of the vortex beams) and also on the azimuthal angle.
In order to quantify this effect we substitute the first non-plane-wave correction (\ref{crossLor}) into the asymmetry formula, Eq.(\ref{Asymm}). 
We arrive at the following compact expression:
\begin{eqnarray}
& \displaystyle
\mathcal A = \langle {\bm b}_{\varphi}\rangle\alpha_{\perp}^{-1}\partial_{\Delta {\bm p}}\zeta_{fi} (s,t) = \text{inv},
\label{Asym}
\end{eqnarray}
or when $\Sigma_x \approx \Sigma_y \approx \Sigma_z \equiv \Sigma$:
\begin{eqnarray}
& \displaystyle
\mathcal A = \frac{2 \Sigma_1^2 \Sigma_2^2}{\Sigma_1^2 + \Sigma_2^2} \left [\frac{\Delta {\bm u}}{|\Delta {\bm u}|} \times \left [\frac{\Delta {\bm u}}{|\Delta {\bm u}|} \times \langle {\bm b}_{\varphi}\rangle \right ]\right ]\cdot \left (\frac{\partial}{\partial {\bm p}_{2}} -\frac{\partial}{\partial {\bm p}_{1}}\right )\zeta_{fi}\Big |_{{\bm p}_{1,2} = \langle {\bm p}\rangle_{1,2}},
\label{Asymm_parax}
\end{eqnarray}
with ${\bm b}_{\varphi}$ from Eq.(\ref{bsubst}) and $\Sigma_{1,2}$ from Eq.(\ref{Sigma}).
This formula could have actually been guessed from the symmetry considerations.
Indeed, for our kinematics the asymmetry can depend only upon the following three vectors:
$\Delta {\bm u}, {\bm b}_{\varphi}, \partial_{\Delta{\bm p}}\zeta_{fi}$,
and, simultaneously, it must be a linear function of the two latter ones. 
The only true scalar that satisfies these criteria is Eq.(\ref{Asymm_parax}).
We would like to stress, however, that this formula was obtained in the lowest order of the perturbation theory in $\alpha^{-1}$ 
and that is why $|\mathcal A| \ll 1$ or, at the best, $|\mathcal A| \lesssim 1$. 
Otherwise these expressions are inapplicable.

Consider a $2\rightarrow 2$ process (not necessarily elastic: say, $pp \rightarrow X,\, ee \rightarrow X$, etc.) with $m_1 = m_2$ in the collider frame of reference (\ref{frame}). 
Using the standard invariant variables,
$$
t = (p_1 - p_3)^2,\ s = (p_1 + p_2)^2,
$$
one can write down the derivative $\partial_{\Delta {\bm p}}$ in this frame as follows
\begin{eqnarray}
& \displaystyle
\frac{\partial}{\partial {\bm p}_{1}} -\frac{\partial}{\partial {\bm p}_{2}} = 8 {\bm p} \frac{\partial}{\partial s} + 4 \left({\bm p}_3 - {\bm p} \right)\frac{\partial}{\partial t}.
\label{der}
\end{eqnarray}
For azimuthally symmetric dispersion with $\Sigma_x \approx \Sigma_y \equiv \Sigma$ and $\alpha^{-1}_{\perp,11} \approx \alpha^{-1}_{\perp,22} \equiv \alpha^{-1}$ in (\ref{alphaperp}),
we get
\begin{eqnarray}
& \displaystyle
\mathcal A = 4\alpha^{-1}\,\langle {\bm b}_{\varphi}\rangle {\bm p}_3\, \frac{\partial\zeta_{fi} (s,t)}{\partial t}.
\label{Asyminframe}
\end{eqnarray}
We shall stick to this model in what follows. When the incoming packets' widths are the same, $\sigma_1 \approx \sigma_2,\, \sigma_{b,1} \approx \sigma_{b,2} \equiv \sigma_b$ and $\Sigma^2 \approx 1/\sigma_{b}^2$ (recal Eq.(\ref{Sigma})),
then
$$
\alpha^{-1} = \frac{2\Sigma_1^2 \Sigma_2^2}{\Sigma_1^2 + \Sigma_2^2} \approx \frac{1}{\sigma_b^2}.
$$
As $\langle {\bm b}_{\varphi}\rangle \equiv \langle {\bm b}_{\varphi}\rangle_{\perp}$, the asymmetry (\ref{Asyminframe}) is odd with respect to 
$$
\phi_3 \rightarrow \phi_3 \pm \pi
$$ 
Therefore, the amplitude's phase $\zeta_{fi}$ \textit{violates an up-down symmetry} in angular distributions of the final particles, if this symmetry takes place without the phase of course.
Note that for the strictly forward scattering, ${\bm p}_3 \rightarrow {\bm p}$, the asymmetry vanishes.

As we have already mentioned in Sec.\ref{Corr}, the averaging of ${\bm b}_{\varphi}$ has appeared because some phases $\varphi_{1,2}$ may not be analytical in the entire ${\bm p}$-domain, 
but contain a finite number of removable singularities. Say, for vortex beams with $\varphi = \ell \phi$ the derivative 
\begin{eqnarray}
& \displaystyle
\frac{\partial \varphi}{\partial {\bm p}} = \ell\, \frac{\hat{{\bm z}} \times {\bm p}}{{\bm p}_{\perp}^2}
\label{der}
\end{eqnarray}
is not analytical for a vanishing transverse momentum. This singularity is removable and the mean value of this,
\begin{eqnarray}
& \displaystyle
\Big \langle\frac{\hat{{\bm z}} \times {\bm p}}{{\bm p}_{\perp}^2} \Big \rangle = \frac{\hat{{\bm z}} \times \langle{\bm p}\rangle}{\langle {\bm p}_{\perp}\rangle^2}\left (1 - e^{-\langle {\bm p}_{\perp} \rangle^2/\sigma^2}\right ) 
,\label{mean_value}
\end{eqnarray}
simply vanishes when $\langle{\bm p}_{\perp} \rangle \rightarrow 0$. 




Note that this asymmetry is a purely quantum effect that vanishes in the plane-wave limit 
and might seem to be counter-intuitive from a classical perspective.
Indeed, for a pair of azimuthally symmetric wave packets their substitution clearly does not alter the (classical) cross section. 
It is violated when either the packets are not-azimuthally symmetric 
(the $2^{\text{nd}}$ scenario) or the particles themselves have some inner structure (atoms, ions, hadrons). 
It is the latter case in which the phase $\zeta_{fi}$ comes into play.

\subsection{$1^{\text{st}}$ scenario: off-center collision of Gaussian beams}

As discussed in Sec.\ref{phases}, there are two ways how one can measure the asymmetry in a collision experiment.
In the first one with two phaseless Gaussian beams collided at a non-vanishing impact-parameter 
one can put $\langle {\bm b}_{\varphi} \rangle = {\bm b} =\{b, 0, 0\}$, where 
$$
b \lesssim \sigma_b, 
$$ 
otherwise the number of events is exponentially suppressed.
Moreover, as clear from Eq.(\ref{Asyminframe}), one should not necessarily swap the beams: 
it is enough to have a non-vanishing $b$ and then to compare angular distributions of the scattered particles 
in the upper- and in the lower semi-spaces. Their difference reveals itself in the asymmetry, which is
\begin{eqnarray}
& \displaystyle
\mathcal A \approx 4\,\frac{p_3}{\sigma_b}\,\sin \theta_{sc}\cos\phi_{sc}\, \frac{\partial\, \zeta_{fi}}{\partial t}.
\label{Eq6}
\end{eqnarray}
It is only linearly attenuated with $\sigma_b$ and its pre-factor has a simple $\sin \theta_{sc}\cos\phi_{sc}$ dependence upon the scattering angles $\theta_{sc} \equiv \theta_3, \phi_{sc} \equiv \phi_3$.
Any deviation of the measured asymmetry from this dependence would be an evidence of a non-trivial phase $\zeta_{fi} (s, t)$.

Further simplifications are possible for elastic scattering in the relativistic case 
with 
\begin{equation}
\displaystyle
p_3 \approx p,\ t \approx - p^2 \theta_{sc}^2,\ \theta_{sc} \ll 1,\ \gamma = \varepsilon/m \gg 1,
\label{elrel}
\end{equation}
and now (compare this with Eq.(\ref{Cijrel}))
\begin{eqnarray}
& \displaystyle
\mathcal A \approx - 2\,\frac{\lambda_c}{\sigma_b}\,\cos\phi_{sc}\,\sqrt{\tau_0}\, \frac{\partial\, \zeta_{fi}}{\partial \tau_0},\ \tau_0 = \frac{-t}{4 m^2},
\label{Eq6t}
\end{eqnarray}
where $\lambda_c = 1/m$ is the Compton wavelength of the incoming particle.
We see that the asymmetry is only linearly attenuated by $\lambda_c/\sigma_b$, 
and it gets bigger when the momentum uncertainty of the beams approaches $m$
and $|t|$ becomes greater than $4 m^2$ (unlike the correction (\ref{Cijrel})). 
As we know, it is exactly when loop diagrams become significant.

Assuming that the phase is a fast function of the scattering angle $\theta_{sc}$, but a slow one of $p$, we get the formula
\begin{equation}
\displaystyle
\mathcal A \approx - 2\,\frac{1}{p \sigma_b}\,\cos\phi_{sc}\, \frac{\partial\, \zeta_{fi}}{\partial \theta_{sc}}, 
\label{Eq6a}
\end{equation}
which shows how the phase changes with the scattering angle. 
Since in our approximation $p \approx \varepsilon = \gamma m$, one can re-write this formula as follows:
\begin{equation}
\displaystyle
\mathcal A \approx - 2\,\frac{\lambda_c}{\sigma_b}\,\cos\phi_{sc}\, \frac{1}{\gamma}\frac{\partial\, \zeta_{fi}}{\partial \theta_{sc}}
 \label{msigma}
\end{equation}
The factors $\lambda_c/\sigma_b$ and $\gamma^{-1} \partial\, \zeta_{fi}/\partial \theta_{sc}$ are Lorentz invariant separately,
and 
for protons the former is of the order of $10^{-10}$ for moderately relativistic beams focused in a spot of $\sim 1 \mu$m 
and of the order of $10^{-8}$ for protons with $p \approx 2$ MeV and focused to $\sigma_b \gtrsim 10$ nm \cite{p, pp}. 
The estimate (\ref{msigma}), however, is inapplicable for such non-relativistic particles.

Conversely, in collision of electrons the ratio $\lambda_c/\sigma_b$ becomes bigger than $10^{-3}$ for $300$-keV electrons focused 
in a spot of the order of $1 \text{\AA}$ \cite{Angstrom} (regardless of the OAM),
although for such intermediate energies the formula (\ref{msigma}) can be used only for qualitative analysis.
According to West and Yennie, for a Coulomb phase on a one-loop level \cite{West}
\begin{eqnarray}
& \displaystyle
\frac{1}{\gamma}\frac{\partial\, \zeta_{fi}}{\partial \theta_{sc}} \sim \frac{\alpha_{em}}{\gamma \theta_{sc}}
\label{zetatheta}
\end{eqnarray}
and hence
\begin{eqnarray}
& \displaystyle
\mathcal A = \mathcal O \left (\frac{\lambda_c}{\sigma_b}\frac{\alpha_{em}}{\gamma\theta_{sc}}\right )
\label{Eq13aaa}
\end{eqnarray}
with $\alpha_{em} \approx 1/137$. This estimate is in accordance with that of the recent paper \cite{I_C}.
In this scenario, we bring two sub-nm-sized electron beams into collision (note that in this case $1/\sigma \sim \sigma_b$), slightly off-center, 
and that is why one ought to be able to control their relative position with the accuracy better than $0.5 \text{\AA}$.
Then angular distributions of the scattered electrons are measured and compared in the upper- and in the lower semi-spaces.
Their difference reveals itself in the asymmetry and its conservative estimate for the scattering angles of $\theta_{sc} \sim 10^{-2}-10^{-1}$ is
\begin{eqnarray}
& \displaystyle
\mathcal |A| \sim 10^{-4} - 10^{-3}
\label{Aestim}
\end{eqnarray}
which is in principle measurable with high statistics. One could further increase it by performing measurements at yet smaller scattering angles 
or by making the impact parameter very large, $b \gg \sigma_b$.
In the latter case, however, the price is a drop in the number of events.

Returning to the elastic scattering of protons, little can be said, unfortunately, in a model-independent way about the factor in the left-hand-side of (\ref{zetatheta}). 
The TOTEM collaboration has managed to perform measurements at the scattering angles lower than $10^{-5}$ at $\sqrt{s} = 8$ TeV \cite{TOTEM_2016}, 
which yields $\gamma \theta_{sc} \sim 10^{-2} - 10^{-1}$, and the hadronic phase $\zeta_{fi}$ itself, unlike the Coulomb one, 
is \textit{not} attenuated by a small parameter $\alpha_{em} \rightarrow \alpha_s$, 
as scattering within a diffraction cone is not described by the perturbation theory. 

\begin{figure}
\center
\includegraphics[width=9.00cm, height=5.50cm]{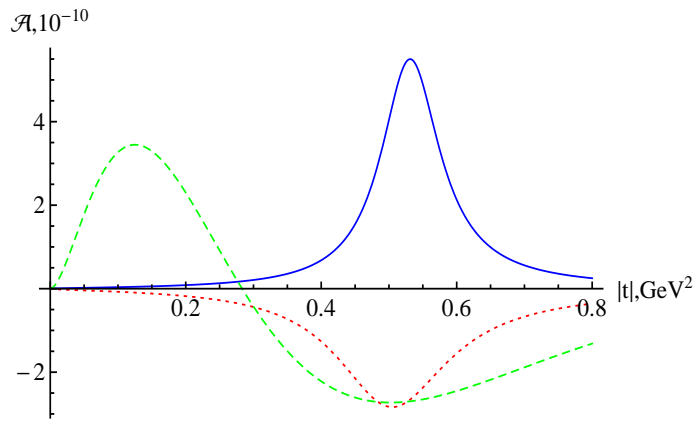}
\caption{The azimuthal asymmetry (\ref{Eq6t}) in proton-proton scattering for different models of the hadronic phase and $\phi_{sc} = 0$. The red dotted line: the standard parametrization, 
the blue solid line: the one by Bailly et al., the green dashed line: peripheral parametrization.
\label{Fig_Asymm}}
\end{figure}

As an example let us take three following models for the hadronic phase employed in the recent experiment at the LHC \cite{TOTEM_2016}:
\begin{eqnarray}
&& \displaystyle
\frac{\partial \zeta_{fi}}{\partial t} = - \frac{\tau}{\tau^2 + (t + |t_0|)^2} - \text{the so-called standard parametrization},
\cr && \displaystyle
\frac{\partial \zeta_{fi}}{\partial t} = - \frac{\rho t_d}{(\rho t_d)^2 + (t - t_d)^2} - \text{the one by Bailly et al. \cite{Bailly}},
\cr && \displaystyle
\frac{\partial \zeta_{fi}}{\partial t} = \zeta_1 (\kappa + \nu t) \left (\frac{-t}{1\, \text{GeV}^2}\right )^{\kappa-1} e^{\nu t} - \text{the so-called} 
\cr && \displaystyle 
\qquad \qquad\qquad\qquad\qquad \qquad\qquad\qquad \text{peripheral parametrization \cite{periph}},
\label{Param}
\end{eqnarray}
where $\rho = \Re M_{fi}/\Im M_{fi} \equiv \rho (t=0)$. 
Taking the same parameters as in \cite{TOTEM_2016}, that is, $\sqrt{s} = 8\, \text{TeV}, \rho = 0.1, t_0 = -0.5\, \text{GeV}^2, \tau = 0.1\, \text{GeV}^2, t_d = -0.53\, \text{GeV}^2, \zeta_1 = 800, \kappa = 2.311, \nu = 8.161\, \text{GeV}^{-2}$, we can estimate the asymmetry in Eq.(\ref{Eq6t}). For the proton beam's width of $\sigma_b \sim 10\, \mu$m we get the results shown in Fig.\ref{Fig_Asymm}.
The derivative of the phase itself is no longer small, but the asymmetry is suppressed by the following factor:
$$
\frac{\lambda_c}{\sigma_b} \sim 10^{-11}.
$$
Although at the small transferred momenta interference of the hadronic phase with the Coulomb one can become prominent \cite{TOTEM_2016},
the asymmetry (unlike the one for electron scattering) stays too small due to the large $\sigma_b$.
Thus the effects of the amplitude's phase are governed by width of the colliding beams.

\subsection{$2^{\text{nd}}$ scenario: colliding beams with phases} 

Within the second scenario, we start with a head-on collision of two vortex beams with the impact-parameter ${\bm b} = 0$, 
the phases $\varphi_{1,2} = \ell_{1,2}\,\phi_{1,2}$, $\ell \equiv \ell_z$, 
and opposite signs of their orbital helicities \cite{I_PRD}. 
The spatial distribution of such beams is no longer Gaussian but a doughnut-shaped one with a minimum on the collision axis 
(see Fig.\ref{Fig1}). As before, we need not necessarily to swap the beams or even change the signs of their OAM.
It is enough to compare angular distributions of the final particles in the upper- and in the lower semi-spaces,
that is, when $\phi_{sc} \rightarrow \phi_{sc} \pm \pi$.

Still working in the frame (\ref{frame}), we find with the help of Eq.(\ref{mean_value}):
\begin{eqnarray}
& \displaystyle
\langle {\bm b}_{\varphi} \rangle = - (\ell_1 + \ell_2)\, \frac{\hat{\bm z} \times \langle{\bm p}\rangle}{\langle{\bm p}\rangle_{\perp}^2}\, (1 - e^{-\langle{\bm p}\rangle_{\perp}^2/\sigma^2}).
\label{Eq7}
\end{eqnarray}
This vector vanishes, together with the asymmetry, when either the total OAM is zero, $\ell_1 + \ell_2 = 0$, 
or one collides the azimuthally symmetric Bessel beams with ${\bm u}_{\perp} = 0$ (which is implied in the frame (\ref{frame})). 
Clearly, what is happening here is that in order to have a non-zero $\mathcal A$ the azimuthal symmetry of the problem must be broken \textit{already in the initial state}, 
exactly as in the previous scenario.

When the impact-parameter is vanishing, violation of the (initial) azimuthal symmetry can be achieved by shifting a phase vortex off the beam's symmetry axis. 
When dealing with the holograms (as in Refs.\cite{Verbeeck, McMorran}), a shift of a fork dislocation off the beam center 
provides a (small) azimuthal asymmetry or, in other words, a non-vanishing transverse momentum (see details, for example, in Ref.\cite{Mair_2001}).
The probability density for such a state is depicted in the right panel of Fig.\ref{Fig1}.
Such a shift is to be small, $\delta\rho \lesssim \sigma_b$, $\delta p_{\perp} \gtrsim 1/\sigma_b,\, \delta\theta \sim 1/(p\sigma_b)$
and it is made to opposite directions for both beams. To put it simply, a non-vanishing transverse momentum plays in this scenario the same role as does a finite impact parameter in the previous section.

By using Eq.(\ref{Asyminframe}), we arrive at the following estimate for the asymmetry: 
\begin{eqnarray}
& \displaystyle
& \displaystyle
\mathcal A \approx -4(\ell_1 + \ell_2) \frac{p_3}{\sigma_b^2 \sigma} \sin \theta_3 \sin (\phi_3 - \phi)\,
\frac{\partial\, \zeta_{fi}}{\partial t}. 
\label{Eq8}
\end{eqnarray}
The major difference between this expression and Eq.(\ref{Eq6}) is appearance of the factor $\ell_1 + \ell_2$, which can be very large.
It might seem therefore that the second scenario with $\ell_{1,2} \gg 1$ provides a much higher value of the asymmetry.
However the price for this increase is again a drop in statistics due to the factor $\exp\{-\ell_{1,2}^2/(2\sigma_{b,1,2}^2{\bm p}_{1,2, \perp}^2)\}$
in the number of events. By analogy with Eq.(\ref{Eq6a}), the factor  
\begin{eqnarray}
& \displaystyle
\frac{\ell_1 + \ell_2}{p \sigma_b^2\sigma}
\label{Eq8a}
\end{eqnarray}
determines sensitivity to the asymmetry in relativistic case. The maximum value of OAM for which the number of events is not suppressed is 
$\ell_{\text{max}} \sim p_{\perp} \sigma_b \sim \sigma \sigma_b$ (see Eq.(\ref{lpackmax})), and that is why 
\begin{eqnarray}
& \displaystyle
\mathcal A \propto \frac{1}{p\sigma_b},
\label{Eq8b}
\end{eqnarray}
exactly as in the previous scenario. 

Since the production of twisted hadrons with azimuthally non-symmetric profiles seems to be more technologically challenging than it is for electrons,
we turn to elastic scattering of the latter particles. In order to maximize the effect, one can take again $300$-keV twisted electrons focused to $\sigma_b \sim 1 \text{\AA}$,
with the monochromaticity\footnote{The beam's monochromaticity in an electron microscope can be as low as $\sigma \sim 1$ eV, however, 
for the electrons focused to a spot of $1 \text{\AA}$ the momentum uncertainty gets higher: $\sigma \lesssim 1/\sigma_b \sim 1$ keV.} $\sigma/p \lesssim 1\%$, and $\sigma \sigma_b \sim \ell_{\text{max}} \sim 1$. 
For measuring the asymmetry, angular distributions of the scattered electrons are to be compared
in the two experiments with $\ell_{1,2} = 1$ and $\ell_{1,2} = -1$, respectively. Alternatively, one can carry out only one experiment with $\ell_{1,2} = 1$ 
when comparing angular distributions in the upper- and in the lower semi-spaces.
The numerical estimate (\ref{Aestim}) stays valid. Since for such a study we need vortex electrons with the azimuthally asymmetric profiles, we would also like to find such states 
for which the requirement of a non-vanishing transverse momentum can be relaxed.

As can be readily seen, it is the case for Airy states as their azimuthal distribution itself is highly asymmetric. 
For collision of two such beams with ${\bm u}_{\perp} = 0$, the phases $\varphi = (\xi_x^3 p_x^3 + \xi_y^3 p_y^3)/3$ (see Eq.(\ref{psiD})), 
and the opposite signs of their parameters ${\bm \xi}_1 = -{\bm \xi}_2 \equiv {\bm \xi} = \{\xi, 0, 0\}$,
we find:
\begin{eqnarray}
&\displaystyle
\langle {\bm b}_{\varphi} \rangle = - \sigma^2 \{\xi^3, 0, 0\},\
\cr 
&\displaystyle \mathcal A \approx -4\, \frac{\sigma^2}{\sigma_b^2}\, \xi^3 p_3 \sin \theta_{sc}\cos \phi_{sc}\, \frac{\partial\, \zeta_{fi} }{\partial t},
\label{Eq10}
\end{eqnarray}
where we have used $\langle p_x^2 \rangle = \langle p_x \rangle^2 + \sigma^2/2$. The typical values of $\xi$ follow from the factor $\exp\{-\Sigma^2(\sigma^2\xi^3/2)^2/2\}$ 
in the probability formula: see Eq.(\ref{Airymax}).
In any case, this yields the same $p_3/\sigma_b$ factor in the asymmetry as in Eq.(\ref{Eq6}) and $\lambda_c/\sigma_b$ for relativistic elastic scattering.
Therefore the use of Airy beams leads to the very same predictions for the asymmetry as in the previous examples.

Moreover, one could think of such a phase $\varphi ({\bm p})$ that maximizes the asymmetry. Within the paraxial case with $\Sigma \ll \langle p \rangle$, however,
the phases are limited by Ineq.(\ref{phaseineq}). That is why the asymmetry stays $\mathcal O (\lambda_c/\sigma_b)$ for all the other types of non-plane-wave states as well.

The idea of using vortex states for probing the amplitude's phase was put forward by Ivanov \cite{I_Phase}. 
By analogy with his work, let us consider now scattering of a light particle by a heave one (say, $e p \rightarrow X, \gamma p\rightarrow X$) with $\sigma_1/\sigma_2 \ll 1$.
Working in a frame in which the longitudinal momentum of the heavy particle is zero, we assume the light one to be in the pure Bessel state with ${\bm u}_{\perp,1} = 0$.
We obtain that the asymmetry,
\begin{eqnarray}
& \displaystyle
\mathcal A 
\propto \ell_2 \sigma_1\,\frac{\sigma_1}{\sigma_2},
\label{Eq13}
\end{eqnarray}
does not depend on the OAM $\ell_1$ of the light particle and, compared to Eq.(\ref{Eq8}), it has an additional small factor $\sigma_1/\sigma_2$, 
which is less than $10^{-3}$ for available beams. 
This factor also appears for the Airy beams when ${\bm p}_1 \ne -{\bm p}_2$ but $\sigma_1 \ll \sigma_2$. 
That is why the higher values of the asymmetry favor the case with $\sigma_1 \sim \sigma_2$, in accordance with the Ref.\cite{I_Phase}.

The difference between the two methods described above can be elucidated by comparing two ways of colliding two rubber balls. 
If the balls are pumped up well, they are azimuthally symmetric and in order to violate this symmetry in scattering we need to collide them slightly off-center.
Conversely, when the balls are deflated they are most likely no longer azimuthally symmetric and that is why they can collide even at a zero impact parameter.
One simply needs to imagine a wave packet with a non-trivial wave front instead of such a deflated ball.

Concluding, scattering experiments probing the Coulomb phase, albeit being on the frontiers of technology, 
can be carried out at the modern electron microscopes, both with the Gaussian beams and with the vortex- and/or Airy ones 
if they are focused to a spot of the order of or less than $1 \text{\AA}$ in diameter.
Predictions for the hadronic (or relative) phase are less encouraging, due to the small ratio $\lambda_c/\sigma_b$, and inevitably model-dependent.

\section{2 packets with phases $\rightarrow$ 2 Bessel states}

Describing detected states as plane waves, we lose information about all the other quantum numbers the evolved state may possess per se.
The simplest example here is a Compton back-scattering of an optical twisted photon by an ultra-relativistic plane-wave electron \cite{Serbo}.
In this case, one may use an orthonormal set of Bessel beams for describing the outgoing particles, which can be treated as entangled in their OAM \cite{Ivanov_PRA_2012}. 

Bearing this in mind, let us choose now two out-states as the pure Bessel ones with the OAM $\ell_3$ and $\ell_4$, and the Wigner functions from Eq.(\ref{WBess}). 
Both the incoming particles are still described with the Gaussian wave packets (\ref{WGaussA}) which can be later generalized to possess complex phases. 
For the sake of simplicity, we shall quantize both the final OAM relative to the same z-axis, which means that the scattering angles should be smaller than the momentum's conical angles. 
Generalization for the case with the so-called orbital helicity \cite{I_S} is straightforward when the final states are also described as wave packets
rather than as the idealized Bessel states. This problem, however, requires tedious calculations and will be tackled elsewhere.
We shall also stick to the model with $\sigma_{ij} = \sigma \delta_{ij}$ throughout this section.

The integration measure for the final states in the probability formula (\ref{M4}) can be chosen as follows (see, for example, \cite{Serbo, I_PRD}):
$$
dn_f = \frac{R}{\pi}\,d\kappa_3\,\frac{L}{2\pi}\,dp_{3,\parallel}\,\frac{R}{\pi}\,d\kappa_4\,\frac{L}{2\pi}\,dp_{4,\parallel}
$$
At this point an important remark is in order. Using the detected non-plane-wave states with a definite set of quantum numbers, 
we imply that there exists an appropriate detector, which is sensitive to this set. For twisted photons, an OAM-sensitive ``detector'' may be thought of as a combination 
of a computer-generated hologram projecting the twisted state back onto the fundamental mode with $\ell = 0$, a mono-mode fiber, 
and a ``usual'' detector -- a setup routinely used in quantum optics with the parametric down-converted twisted photons (see, for example, \cite{Mair_2001, F-A_2012}).
For electrons and other massive vortex particles, an analogous registration scheme may also include a pair of appropriate holograms and a CCD camera.

Let us now derive a probability formula analogous to Eqs.(\ref{MApp}),(\ref{Eq6Alt}) but with the final Bessel states. 
First we find the following relation for the Wigner function (\ref{WBess}):
\begin{eqnarray}
& \displaystyle
\int d^3R\, e^{i{\bm k}{\bm R}}\, n ({\bm r} + {\bm R}, {\bm p}, t; \ell) = \frac{(2\pi)^4}{R L}\, \delta (p_z - p_{\parallel})\,\frac{\Theta (\kappa - p_{\perp})}{p_{\perp} \sin \xi} 
\cr & \displaystyle 
\times \Big (\delta^{(3)} \left({\bm k} - 2\tan \xi\, [\hat{{\bm z}} \times {\bm p}]\right )\, \exp \Big\{2i \Big (\tan \xi\, [{\bm r} \times {\bm p}]_z - \ell \xi\Big )\Big\} + 
\cr & \displaystyle 
+ \, \delta^{(3)} \left({\bm k} + 2\tan \xi\, [\hat{{\bm z}} \times {\bm p}]\right )\, \exp \Big\{-2i \Big (\tan \xi\, [{\bm r} \times {\bm p}]_z - \ell \xi\Big )\Big\}\Big ).
\label{RBess}
\end{eqnarray}
where $\sin \xi = \sqrt{1 - (p_{\perp}/\kappa)^2}$. This yields the following formula for the correlator from Eq.(\ref{M4}):
\begin{eqnarray}
& \displaystyle
\mathcal L^{(4)} = \upsilon \frac{(2\pi)^{12}}{(RL)^2} \Big (\frac{2}{\sigma_1 \sigma_2}\Big)^3\, \delta (p_{3, z} - p_{3, \parallel})\, \delta (p_{4, z} - p_{4, \parallel})\,
\delta \left ({\bm k}_1 {\bm u}_1 + {\bm k}_2 {\bm u}_2 \right ) \times
\cr & \displaystyle 
\frac{\Theta (\kappa_3 - p_{3, \perp}) \Theta (\kappa_4 - p_{4, \perp})}{p_{3, \perp} p_{4, \perp} \sin \xi_3 \sin \xi_4}\, \exp \Big\{-i{\bm r}_{1, 0}{\bm k}_1 - i{\bm r}_{2, 0}{\bm k}_2 - \frac{{\bm k}_1^2}{(2\sigma_1)^2} - \frac{{\bm k}_2^2}{(2\sigma_2)^2} - 
\cr & \displaystyle 
- \frac{({\bm p}_1 - \langle {\bm p}\rangle_1)^2}{\sigma_1^2} - \frac{({\bm p}_2 - \langle {\bm p}\rangle_2)^2}{\sigma_2^2}\Big\}
\Big (F(\xi_3, \xi_4) + F(-\xi_3, \xi_4) + F(\xi_3, -\xi_4) + F(-\xi_3, -\xi_4)\Big ),
\cr & \displaystyle 
F(\xi_3, \xi_4) = \delta^{(3)} ({\bm k}_3 - 2\tan \xi_3\, [\hat{\bm z} \times {\bm p}_3])\, \delta^{(3)} ({\bm k}_4 - 2\tan \xi_4\, [\hat{\bm z} \times {\bm p}_4])\, e^{-2i (\xi_3 \ell_3 + \xi_4 \ell_4)}
\label{corr}
\end{eqnarray}
where we have used ${\bm k}_1 + {\bm k}_2 - {\bm k}_3 - {\bm k}_4 = 0$. 

Now we return to Eq.(\ref{M4}), integrate over ${\bm k}_3$ and ${\bm k}_4$, and then, similar to the procedure employed in the previous sections, we make an expansion of $d \sigma \left ({\bm k}, {\bm p}\right )$
over the small ${\bm k}_1, {\bm k}_2$. The integral over ${\bm k}_2$ is then eliminated with the use of the delta-function, $\delta ({\bm k}_1 + {\bm k}_2 - {\bm k}_3 - {\bm k}_4)$, 
and for the $k_1$-dependent part we get the integral similar to (\ref{k}) with the same matrix $B_{ij}$, but this time with 
\begin{eqnarray}
& \displaystyle
{\bm A} = i ({\bm b} + t^{\prime} \Delta {\bm u}) - \alpha_2 {\bm k}_f,\ \alpha_2 = \frac{1}{2\sigma_2^2} - \frac{it}{4\varepsilon_2},
\cr & \displaystyle 
{\bm k}_f = {\bm k}_3 + {\bm k}_4 \equiv {\bm k}_f (\pm \xi_3, \pm \xi_4) = \pm 2\tan \xi_3 [\hat{\bm z} \times {\bm p}_3] \pm 2\tan \xi_4 [\hat{\bm z} \times {\bm p}_4],
\label{ABess}
\end{eqnarray}
where ${\bm k}_f (\xi_3, \xi_4)$ is different for all four summands in (\ref{corr}), and ${\bm b}$ is the impact-parameter between the two wave-packets. 
The integral over $t^{\prime}$ is again gaussian, and the final result for the probability represents a sum of four terms:
\begin{eqnarray}
& \displaystyle
dW = dW [\xi_3, \xi_4] + dW [-\xi_3, \xi_4] + dW [\xi_3,-\xi_4] + dW [-\xi_3,-\xi_4],
\cr & \displaystyle 
dW [\xi_3, \xi_4] = d\kappa_3\,d\kappa_4\,dp_{3,\parallel}\,dp_{4,\parallel}\, \left(\frac{2}{\sigma_1 \sigma_2}\right)^3 \frac{(2\pi)^3}{\pi^2} \int \prod\limits_{i=1}^{4}\frac{d^3 p_i}{(2\pi)^3} \frac{dt}{2\pi}\, 
\cr & \displaystyle 
\times (2\pi)^3 \delta^{(3)} ({\bm p}_1 + {\bm p}_2 - {\bm p}_3 - {\bm p}_4)\delta (p_{3, z} - p_{3, \parallel})\, \delta (p_{4, z} - p_{4, \parallel})\, \frac{\Theta (\kappa_3 - p_{3, \perp}) \Theta (\kappa_4 - p_{4, \perp})}{p_{3, \perp} p_{4, \perp} |\sin \xi_3| |\sin \xi_4|} 
\cr & \displaystyle 
\times\frac{1}{\sqrt{\det B\, \Delta u B^{-1} \Delta u}}\, \exp\Big\{it \left (\varepsilon_1 ({\bm p}_1) + \varepsilon_2 ({\bm p}_2) - \varepsilon_3 ({\bm P}_{3,+}) - \varepsilon_4 ({\bm P}_{4,+})\right ) -
\cr & \displaystyle 
- \frac{({\bm p}_1 - \langle{\bm p}\rangle_1)^2}{\sigma_1^2} -\frac{({\bm p}_2 - \langle{\bm p}\rangle_2)^2}{\sigma_2^2}- \frac{\alpha_2}{2} {\bm k}_f^2 - i {\bm r}_{0, 2} {\bm k}_f  - 2i (\ell_3 \xi_3 + \ell_4 \xi_4) - 
\cr & \displaystyle 
- \frac{1}{2} (b + i\alpha_2 k_f) B^{-1} (b + i\alpha_2 k_f) - \frac{1}{2} \frac{1}{\Delta u B^{-1} \Delta u}\, \left (\alpha_2\, \Delta u B^{-1} k_f + {\bm u}_2{\bm k}_f -i \Delta u B^{-1} b \right )^2\Big\}
\cr & \displaystyle
 \times \Big \{T_{fi} ({\bm p}_1, {\bm p}_2, {\bm P}_{3,+}, {\bm P}_{4,+}) T_{fi}^* ({\bm p}_1, {\bm p}_2, {\bm P}_{3,-}, {\bm P}_{4,-}) + \mathcal O (\sigma^2)\Big \}
\label{MAppBess}
\end{eqnarray}
where 
$$
{\bm P}_{3,\pm} = {\bm p}_3 \pm \tan \xi_3\, [\hat{\bm z} \times {\bm p}_3],\, {\bm P}_{4,\pm} = {\bm p}_4 \pm \tan \xi_4\, [\hat{\bm z} \times {\bm p}_4].
$$
As before, when the incoming particles possess phases we need to make the substitution (\ref{bsubst}) and also 
$$
{\bm r}_{0, 2} \rightarrow {\bm r}_{0, 2} - \partial \varphi_2 ({\bm p}_2)/\partial {\bm p}_2.
$$

Compared to Eq.(\ref{MApp}), this probability formula reveals several new features:
\begin{itemize}
\item
It depends not only on the effective impact parameter ${\bm b}_{\varphi}$, but also on the initial condition ${\bm r}_{0,2}$ by itself;
\item
The function in the exponent is no longer ${\bm b} \rightarrow - {\bm b}$ symmetric
;
\item
Even in the lowest $\sigma$-order, there is a contribution from the phase of the scattering amplitude. 
\end{itemize}
It must be noted, however, that the last effect takes place solely because of the non-normalizable nature of pure Bessel beams.
If we had used the well-normalized wave packets with the OAM (\ref{WGaussOAM}) instead, we would have come to $|T_{fi} ({\bm p})|^2$
in the leading order, as in Eq.(\ref{MApp}).

If we sum over all the final OAM, the resultant expression no longer has these features and looks very similar to Eq.(\ref{MApp}):
\begin{eqnarray}
& \displaystyle
\sum\limits_{\ell_3, \ell_4} dW = d\kappa_3\,d\kappa_4\,dp_{3,\parallel}\,dp_{4,\parallel}\, \left(\frac{2}{\sigma_1 \sigma_2}\right)^3 \frac{(2\pi)^5}{\pi^2} \int \prod\limits_{i=1}^{4}\frac{d^3 p_i}{(2\pi)^3} \frac{dt}{2\pi}\, (2\pi)^3 \delta^{(3)} ({\bm p}_1 + {\bm p}_2 - {\bm p}_3 - {\bm p}_4)
\cr & \displaystyle 
\times \delta (p_{3, z} - p_{3, \parallel})\, \delta (p_{4, z} - p_{4, \parallel})\, \delta (\xi_3) \delta (\xi_4) \frac{\Theta (\kappa_3 - p_{3, \perp}) \Theta (\kappa_4 - p_{4, \perp})}{p_{3, \perp} p_{4, \perp} |\sin \xi_3| |\sin \xi_4|} \frac{1}{\sqrt{\det B\, \Delta u B^{-1} \Delta u}}
\cr & \displaystyle 
\times\, \exp\Big\{it \left (\varepsilon_1 ({\bm p}_1) + \varepsilon_2 ({\bm p}_2) - \varepsilon_3 ({\bm p}_3) - \varepsilon_4 ({\bm p}_4)\right )
- \frac{({\bm p}_1 - \langle{\bm p}\rangle_1)^2}{\sigma_1^2} -\frac{({\bm p}_2 - \langle{\bm p}\rangle_2)^2}{\sigma_2^2}
\cr & \displaystyle  
- \frac{1}{2}\, b B^{-1} b + \frac{1}{2} \frac{\left (\Delta u B^{-1} b \right )^2}{\Delta u B^{-1} \Delta u}\Big\}
\Big \{|T_{fi} ({\bm p})|^2 + \mathcal O (\sigma^2)\Big \}
\label{MAppBessSumm}
\end{eqnarray}
The integrals over $\xi_{3,4}$ can be evaluated with the use of the following identity (see Eq.(\ref{xi})):
$$
\int\limits_0^{\infty} dp_{\perp} \Theta (\kappa - p_{\perp}) = \int\limits_{0}^{\pi/2} d\xi\, \kappa \sin \xi,\,\, \text{and so}\,\, \int\limits_0^{\infty} dp_{\perp} \frac{\Theta (\kappa - p_{\perp})}{p_{\perp} \sin \xi}\, \delta (\xi) = 1. 
$$

As can be readily checked, the final (evolved) state of two vortex particles is non-separable,
$$
|\ell_3, \ell_4\rangle \ne \ |\ell_3 \rangle | \ell_4 \rangle,
$$
and therefore OAM-entangled. Indeed, in the idealized transition from a two-particle Bessel state $|\ell_{\text{in}}\rangle$ to $|\ell_{\text{out}}\rangle$, 
the evolved one reads:
\begin{eqnarray}
& \displaystyle
|\ell_{\text{out}}\rangle = \hat{S}\, |\ell_{\text{in}}\rangle = \sum\limits_{m, n} S_{\{m,n\}, \ell_{\text{in}}}\, |m, n \rangle,\,\, \text{where}\,\,\, S_{\{m,n\},\ell_{\text{in}}} \propto \,\,
\delta (\varepsilon_{\text{in}} - \varepsilon_m - \varepsilon_n)\cr 
& \displaystyle \times \delta (p_{z, \text{in}} - p_{z, m} - p_{z, n}) \delta (\kappa_{\text{in}} - \kappa_{m} - \kappa_n)\,
\delta_{\ell_{\text{in}},m+n}\, T_{\{m,n\},\ell_{\text{in}}},
\label{ellout}
\end{eqnarray}
where $T_{\{m,n\},\ell_{\text{in}}}$ is the scattering amplitude. These states obviously do not factorize,
\begin{eqnarray}
& \displaystyle
|\ell_{\text{out}}\rangle \propto \sum\limits_{n} T_{\{m = \ell_{\text{in}} - n,n\},\ell_{\text{in}}}\, |m = \ell_{\text{in}} - n\rangle\, | n \rangle.
\label{ellout2}
\end{eqnarray}
An alternative criterion of entanglement is non-factorization of the probability (\ref{MAppBess}) (see, for example, Refs.\cite{Cirone_2005, F-A_2012}):
\begin{eqnarray}
& \displaystyle W (\ell_3, \ell_4) \ne W (\ell_3) W (\ell_4),\label{En8}
\end{eqnarray} 
where $W (\ell_3) = \sum_{\ell_4} W (\ell_3, \ell_4), W (\ell_4) = \sum_{\ell_3} W (\ell_3, \ell_4)$. 
A quantitative estimate of the OAM entanglement (or degree thereof) can be obtained
by using the following entanglement measure \cite{Cirone_2005}:
\begin{eqnarray}
& \displaystyle \mathcal E = \frac{1}{2}\sum\limits_{\ell_3,\ell_4}\left |\left (W (\ell_3) W (\ell_4) - W (\ell_3, \ell_4)\right )\right |,\label{En8}
\end{eqnarray}
which varies from $0$ (no entanglement) to $1$ (maximum entanglement) and it is obviously finite in our case. 

Note that the final Bessel states are monochromatic: they posses definite energy but undetermined azimuthal component of the momentum $p_{\phi}$, 
according to the angular momentum-angle uncertainty relations \cite{OAM_uncertainty, Barnett_1990, F-A_2004, F-A_2011}. The OAM-entanglement appears because the azimuthal components 
of the wave functions do not factorize in $|\ell_3, \ell_4\rangle$, that is, as a result of $p_{\phi}$-interference. 
We would like to emphasize, however, that this happens only when both the incoming states are described with the well-normalized wave packets.
Indeed, even if the in-states have no OAM whatsoever, their \textit{OAM spectra} are finite (see, for example, \cite{Barnett, PRA})
and, as a result, there is some overlap between both the azimuthal distributions. It is precisely this overlap that makes the final state a coherent superposition of the one-particle ones 
and the entanglement measure (\ref{En8}) finite.




\section{Summary and outlook}

We have developed the relativistic scattering theory beyond the plane-wave approximation in the paraxial regime when the incoming packets are narrow in the momentum space.
The Wigner formalism turns out, therefore, to be the rather powerful tool that allows one to study effects accessible neither in the plane-wave approximation nor in the quasi-classical regime.
These non-plane-wave effects are brought about by a finite overlap of the incoming wave packets.
Depending on the phases, the packets represent the coherent states, the vortex beams carrying orbital angular momentum, the Airy beams, as well as their various generalizations.
We have derived the general model-independent expressions for the probability and for the cross section and, when the non-plane-wave effects are small,
have also obtained the analytical formulas for the first corrections to the plane-wave results.

In the latter case, along with the ``kinematic'' terms, $\lambda_c^2/\sigma_b^2$, there are also corrections that depend on the amplitude's model,
that is, $d\sigma^{(1)} \propto f (s,t)\, \lambda_c^2/\sigma_b^2$. In a region of parameters where the function $f$ is large,
this correction is no longer vanishing and it can reach the values up to about $10-20\%$. 
For scattering of electrons with intermediate energies in QED, this happens at the scattering angles of a few tenths of a degree 
and at yet smaller angles for relativistic case.

Perhaps the most compelling finding of this study appears to be the azimuthal up-down asymmetry brought about by the scattering amplitude's phase.
It is only linearly attenuated by the small parameter $\lambda_c/\sigma_b$. We have discussed two methods for probing 
this effect in experiments either with the conventional Gaussian beams or with such novel states as the vortex particles and the Airy beams.
For Coulomb phase, the asymmetry is bigger than $10^{-3}-10^{-4}$ for beams of the modern electron microscopes with the energies of hundreds of keV or less.
For hadronic phase in proton-proton collisions, the similar effects are much weaker due to the ratio $\lambda_c/\sigma_b \sim 10^{-11}$ for beams at the LHC.

The Wigner formalism is thus alternative and complementary to such well-developed quasi-classical methods 
as the trajectory-coherent approach with $\hbar$ being the small parameter \cite{Bagrov_TCS} 
or the operator method in which the ultrarelativistic motion is also implied \cite{BLP}.
Neither of these approaches has an advantage of explicit Lorentz invariance,
although covariant generalization of the Gaussian wave packets seems to be feasible \cite{Packets, Packets_2}.

As we have also demonstrated, in elastic scattering of two particles at least one of which carries OAM the final pair gets OAM-entangled.
In addition to the standard optical technique of the parametric down conversion, such a scattering (or annihilation) could become another method 
for obtaining the OAM-entangled beams, not only of photons but of the massive particles (including hadrons) as well.
The somewhat similar spin-entanglement of final electrons has been recently studied experimentally in \cite{Art}.
The analogous procedure can also be applied to other non-plane-wave beams with other sets of quantum numbers.
It is of general interest, therefore, to generalize these results when the final states also represent wave packets with phases.
This implies that they are detected with the appropriate apparatus and localized spatially as well as temporarily.




I am grateful to E.\,Akhmedov, V.\,Bagrov, I.\,Ginzburg, I.\,Ivanov, P.\,Kazinski, G.\,Kotkin, V.\,Serbo, O.\,Skoromnik, A.\,Zhevlakov and, especially, to A.\,Di Piazza 
for many fruitful discussions and criticism. I also would like to thank C.\,H.\,Keitel, A.\,Di Piazza and S.\,Babacan for their hospitality 
during my stay at the Max-Planck-Institute for Nuclear Physics in Heidelberg. 
This work is supported by the Alexander von Humboldt Foundation (Germany) and by the Competitiveness Improvement Program of the Tomsk State University.

\end{document}